\documentclass[twocolumn]{aastex631}
\usepackage{amsmath}
\usepackage{longtable}

\usepackage{array}
\newcolumntype{C}[1]{>{\centering\arraybackslash}p{#1}}
\usepackage{footnote}
\makesavenoteenv{tabular}
\usepackage{graphicx}
\usepackage{tablefootnote}

\usepackage{xcolor}

\begin{document}

\title{A neutral hydrogen absorption study of cold gas in the outskirts of the Magellanic Clouds using 
the  GASKAP-HI survey}

\correspondingauthor{Hongxing Chen}
\email{hchen792@wisc.edu}

\author[0009-0005-1781-5665]{Hongxing Chen}
\affiliation{University of Wisconsin–Madison, Department of Astronomy, 475 N Charter St, Madison, WI 53703, USA}

\author[0000-0002-3418-7817]{Sne{\v z}ana Stanimirovi{\'c}}
\affiliation{University of Wisconsin–Madison, Department of Astronomy, 475 N Charter St, Madison, WI 53703, USA}

\author[0000-0001-9504-7386]{Nickolas~M.~Pingel}
\affiliation{University of Wisconsin–Madison, Department of Astronomy, 475 N Charter St, Madison, WI 53703, USA}

\author[0000-0002-4899-4169]{James Dempsey}
\affiliation{Research School of Astronomy \& Astrophysics, The Australian National University, Canberra ACT 2611, Australia}
\affiliation{CSIRO Information Management and Technology, GPO Box 1700 Canberra, ACT 2601, Australia}

\author[0000-0003-4213-8094]{Frances Buckland-Willis}
\affiliation{Laboratoire de Physique de l’Ecole Normale Sup\'erieure, ENS, Universit\'e PSL, CNRS, Sorbonne Universit\'e, Universit\'e de Paris, 24 rue Lhomond, 75005 Paris Cedex 05, France}

\author[0000-0002-7633-3376]{Susan E. Clark}
\affiliation{Department of Physics, Stanford University, Stanford, CA 94305, USA}
\affiliation{Kavli Institute for Particle Astrophysics \& Cosmology, P.O. Box 2450, Stanford University, Stanford, CA 94305, USA}

\author[0000-0002-9214-8613]{Helga Dénes}
\affiliation{School of Physical Sciences and Nanotechnology, Yachay Tech University, Hacienda San José S/N, 100119, Urcuquí, Ecuador}

\author[0000-0002-6300-7459]{John M. Dickey}
\affiliation{School of Natural Sciences, Private Bag 37, University of Tasmania, Hobart, TAS, 7001, Australia}

\author[0000-0002-1495-760X]{Steven Gibson}
\affiliation{Department of Physics and Astronomy, Western Kentucky University, Bowling Green, KY 42101, USA}

\author[0000-0001-7105-0994]{Katherine Jameson
}
\affiliation{Caltech Owens Valley Radio Observatory, Pasadena, CA 91125, USA}

\author[0000-0002-6637-9987]{Ian Kemp}
\affiliation{International Centre for Radio Astronomy Research (ICRAR), Curtin University, Bentley, WA 6102, Australia}

\author[0000-0002-4814-958X]{Denis Leahy}
\affiliation{Department of Physics \& Astronomy, University of Calgary, Calgary, AB T2N 1N4, Canada}

\author[0000-0002-9888-0784]{Min-Young Lee}
\affiliation{Korea Astronomy and Space Science Institute, 776 Daedeok-daero, Daejeon 34055, Republic of Korea}
\affiliation{Department of Astronomy and Space Science, University of Science and Technology, 217 Gajeong-ro, Daejeon 34113, Republic of Korea}

\author[0000-0001-6846-5347]{Callum Lynn}
\affiliation{Research School of Astronomy \& Astrophysics, The Australian National University, Canberra ACT 2611, Australia}

\author[0000-0003-0742-2006]{Yik Ki Ma}
\affiliation{Max-Planck-Institut f\"ur Radioastronomie, Auf dem H\"ugel 69, 53121 Bonn, Germany}
\affiliation{Research School of Astronomy \& Astrophysics, The Australian National University, Canberra ACT 2611, Australia}

\author[0000-0003-2730-957X]{N. M. McClure-Griffiths}
\affiliation{Research School of Astronomy \& Astrophysics, The Australian National University, Canberra ACT 2611, Australia}

\author[0000-0002-7743-8129]{Claire E. Murray}
\affiliation{Space Telescope Science Institute, 3700 San Martin Drive, Baltimore, MD 21218, USA}
\affiliation{Department of Physics \& Astronomy, Johns Hopkins University, 3400 N. Charles Street, Baltimore, MD 21218, USA}

\author[0000-0002-2712-4156]{Hiep Nguyen}
\affiliation{Research School of Astronomy \& Astrophysics, The Australian National University, Canberra ACT 2611, Australia}

\author[0000-0002-2082-1370]{Lucero Uscanga}
\affiliation{Departamento de Astronom\'ia, Universidad de Guanajuato, A.P. 144, 36000 Guanajuato, Gto., Mexico}

\author[0000-0002-1272-3017]{Jacco Th. van Loon}
\affiliation{Lennard-Jones Laboratories, Keele University, ST5 5BG, UK}

\author[0000-0002-1424-3543]{Enrique V\'azquez-Semadeni}
\affiliation{Instituto de Radioastronom\'ia y Astrof\'isica, Universidad Nacional Aut\'onoma de M\'exico, Apdo. Postal 3-72, Morelia, Michoac\'an, M\'exico}

\begin{abstract}

Cold neutral hydrogen (HI) is a crucial precursor for molecular gas formation and can be studied via HI absorption. 
This study investigates HI absorption in low column density regions of the Small and Large Magellanic Clouds (SMC and LMC) using the Galactic-ASKAP HI (GASKAP-HI) survey, conducted by the Australian Square Kilometer Array Pathfinder (ASKAP). 
We select 10 SMC directions in the outer regions and 18 LMC directions, with 4 in the outskirts and 14 within the main disk.
Using the radiative transfer method, we decompose the emission and absorption spectra into individual cold neutral medium (CNM) and warm neutral medium (WNM) components. 
In the SMC, we find HI peak optical depths of 0.09 -- 1.16, spin temperatures of $\sim$ 20 -- 50 K, and CNM fractions of 1 -- 11\%. In the LMC, optical depths range from 0.03 to 3.55, spin temperatures from $\sim$ 10 to 100 K, and CNM fractions from 1\% to 100\%. 
The SMC's low CNM fractions likely result from its low metallicity and large line-of-sight depth. Additionally, the SMC's outskirts show lower CNM fractions than the main body, potentially due to increased CNM evaporation influenced by the hot Magellanic Corona. Shell motions dominate the kinematics of the majority of CNM clouds in this study and likely supply cold HI to the Magellanic Stream. In the LMC, high CNM fraction clouds are found near supergiant shells, where thermal instability induced by stellar feedback promotes WNM-to-CNM transition. Although no carbon monoxide (CO) has been detected, enhanced dust shielding in these areas helps maintain the cold HI.


\end{abstract}


\section{Introduction} \label{sec:intro}

Neutral atomic hydrogen (HI), a fundamental building block of the Universe, plays a critical role in the evolution of galaxies. Star formation within galaxies involves a dynamic recycling of material within the interstellar medium (ISM), with HI being central to this process. As the most abundant neutral gas in galaxies \citep{Carilli2013}, HI is essential in fueling star formation as it regulates the galaxy's ability to form molecular gas \citep{Clark2012,Klessen2016}. 
The distribution of HI, varying with temperature and density, reflects the complex radiative and dynamic processes within the ISM. 

HI exists in multiple phases in the ISM. The two long-lived phases are the warm neutral medium (WNM) and cold neutral medium (CNM) \citep{McKee1977}. The CNM, as observed in the Milky Way \citep{Wolfire2003}, typically has kinetic temperatures $T_k \sim 40 - 200$ K and volume densities $n \sim 10 - 100$ cm$^{-3}$ and the WNM has  $T_k \sim 4000 -8000$ K and $n \sim 10^{-2} - 1$ cm$^{-3}$. 
The amount of HI in the CNM phase is important for driving star formation as molecular hydrogen is largely formed out of the CNM \citep{Krumholz2009,Kennicutt2012,Bialy2019}. The coexistence of the WNM and CNM is governed by a delicate balance of heating and cooling mechanisms and is set within a dynamic equilibrium of pressures. 
This equilibrium, as indicated by theoretical models and numerical simulations, is significantly influenced by key physical properties such as the ambient interstellar radiation field, metallicity, dust properties, and interstellar turbulence. These processes can affect the heating/cooling rates, the molecule formation and destruction rates, and the resultant chemical and thermal state of the ISM \citep{Field1969,Wolfire1995, Wolfire2003, Liszt2002, Glover2012, Glover2014, Bialy2019}.

Metallicity plays an important role in shaping the properties of HI phases with theoretical studies suggesting that the CNM should be colder and less abundant in low metallicity galaxies, e.g. \cite{Bialy2019}. 
While the dependence of cooling and
heating on metallicity largely cancels out (within the metallicity range explored here), the
CNM fraction depends subtly on the metallicity via UV radiation transfer and grain charging.
Recent numerical simulations of the star-forming ISM, including an explicit UV radiation
transfer and photochemistry \citep{Kim2023}, show that the reduced dust attenuation of FUV
radiation at low metallicity makes photoelectric heating more efficient \citep{Kim2024}. The
net effect is more reduction in cooling than heating, and as a consequence, a decrease in the CNM fraction.
Additionally, the scarcity of dust at low metallicity results in lower abundances of molecular hydrogen ($\text{H}_2$) and carbon monoxide (CO), compared to conditions in higher metallicity galaxies \citep{Dobbs2014}. Metallicity profoundly impacts the structure and dynamics of molecular clouds, thereby directly influencing the star formation processes \citep{Rubio1993, Bolatto2008, Heyer2009, Hughes2010, Schruba2012}.

The Magellanic Clouds -- including both the Large and Small Magellanic Cloud (LMC and SMC) -- provide a perfect place to study the properties of the CNM and WNM in low metallicity galaxies, which are similar to early galactic environments. The LMC, an irregular low mass galaxy with a metallicity of 0.5 $Z_\odot$ \citep{Olszewski1991,deGrijs2014}, is the nearest gas-rich galaxy viewed almost face-on and at a distance of 50 kpc. Similarly, the SMC, another part of the Magellanic System, exhibits a  lower metallicity of 0.2 $Z_\odot$ \citep{Dufour1984} at a distance of 60 kpc \citep{deGrijs2015}.

The 21-cm wavelength emission arising from the hyperfine spin-flip transition of HI provides a practical method for probing the HI content of galaxies using radio telescopes. This approach, particularly through the observation of HI absorption against background radio continuum sources, is essential for directly tracing cold and/or optically thick HI in galaxies \citep{Heiles2003, Murray2015}. 
The WNM, on the other hand, often requires high sensitivity of the HI absorption spectra, thus most HI absorption detections are sensitive to the CNM, and occasionally thermally unstable WNM \citep{McClure-Griffiths2023}. By comparing HI absorption with nearby HI emission, it becomes possible to measure the excitation temperature (or spin temperature, $T_s$) and column density ($N_{\mathrm{HI}}$) of CNM and WNM components along the observed line-of-sight (LOS).  

Previous studies of the cold HI in the SMC by \cite{Dickey2000} used the Australia Telescope Compact Array (ATCA) to examine HI absorption in the direction of 32 continuum sources and found that the spin temperature of the CNM clouds was generally lower than what was found in the Milky Way. They suggested that HI absorption likely originates from photodissociated $\text{H}_2$ in cold clouds, resulting in temperatures so low that these regions are mostly molecular in the Milky Way. Subsequent research by \cite{Jameson2019}, using the enhanced sensitivity and velocity resolution of the ATCA, investigated 55 continuum sources behind the SMC. They observed a typical spin temperature of CNM clouds at $\sim$ 30 K, which is lower than  $\sim$ $50-70$ K found in the Milky Way \citep{Heiles2003,Murray2018}. Additionally, they reported an average cold gas fraction of $\sim$ 20\%, which is similar to the low end of the CNM fraction distribution observed in the Milky Way \citep[e.g. mostly in the $10-80$\% range][]{Heiles2003,Murray2018}. 

For the LMC, initial HI absorption surveys by \cite{Dickey1994} and \cite{Marx-Zimmer2000} utilized ATCA to investigate more than 50 background continuum sources. \cite{Dickey1994} proposed that the LMC either has a high CNM fraction or has a very low CNM temperature. \cite{Marx-Zimmer2000} reported a CNM temperature of $\sim$ 30 K in the vicinity of 30 Dor, the region with the highest HI column density in the LMC. \cite{Liu2021} later conducted a more sensitive survey towards 92 continuum sources across the LMC using the ATCA. They found a value of $\sim$ 30 K for the individual CNM components, similar to what was found for the SMC. The average cold gas fraction was estimated as $\sim$ 14\%, which is surprisingly lower than that of the SMC.

The recent Galactic-ASKAP (GASKAP) survey, utilizing the Australian Square Kilometre Array Pathfinder (ASKAP) radio telescope, brings unprecedented resolution and a wide field-of-view of the Magellanic Clouds. \cite{Dempsey2022} presented a pilot GASKAP-HI survey focusing on the SMC, which provided absorption spectra towards 229 continuum sources, substantially increasing the number of sources and the quality of the spectra compared to the previous survey. They found a lower cold gas fraction (11\%) than \cite{Jameson2019} and suggested that this is because the GASKAP survey provides a more comprehensive representation of the SMC's ISM. The global CNM fractions of the SMC and LMC are now in line with expectations based on metallicity differences, see \cite{Stanimirovic2024}.

The GASKAP survey provided the largest number of  HI absorption-detected sources in both the LMC and SMC, making it possible to contrast the cold gas properties in the low vs. high column density regions and search for cold HI in the outskirts of the Clouds.
In the Milky Way, the CNM fraction decreases from $\sim 60$\% to 
few \% over the column density range from $\sim 10^{21}$ $\text{cm}^{-2}$ to $10^{19} $ $\text{cm}^{-2}$ \citep{Heiles2003,Stanimirovic2005,Murray2018}. Due to the lack of large samples of absorption spectra, regional variations of cold HI properties in the Magellanic Clouds have been impossible to investigate.
In addition, low column density regions located primarily in the outskirts of the SMC and LMC, offer valuable insights into the existence and origins of cold gas in the Magellanic Stream. Recently, \cite{Dempsey2020} suggested that fragments of shells ejected from the SMC could supply cold gas to the Magellanic Stream, though their conclusion was based only on a single source. With the expanded sample from the GASKAP survey, we now have more sources to test this hypothesis.

The primary goal of this study is to investigate the properties of cold HI gas in the outskirts of the SMC and LMC, such as spin temperature and the CNM fraction. 
While studies by \cite{Jameson2019} and \cite{Dempsey2022} in the SMC as well as \cite{Liu2021} in the LMC analyzed samples of HI absorption spectra, their samples were smaller and of 
lower sensitivity relative to GASKAP's spectra.
As shown in \cite{Dempsey2022}, the GASKAP survey provides a large sample of high-quality HI absorption spectra with a more uniform sensitivity relative to previous studies. While \cite{Dempsey2022} analyzed HI absorption spectra for the entire SMC using integrated properties, here we 
use the Gaussian decomposition method with radiative transfer \citep[developed by][]{Heiles2003} to fully decompose spectra into the WNM and CNM components, providing more accurate estimates of the CNM fraction. This allows us to explore the spatial distributions of cold HI properties, examine their surrounding environments, and investigate their potential connection to the extended regions of the Magellanic Stream.

This paper is organized as follows. Section~\ref{sec:obs} describes the observations and the extraction of HI absorption spectra in the direction of background radio continuum sources. Section~\ref{sec:source_sel} explains how we select target background sources, as well as the measurements of the corresponding HI emission spectra. Section~\ref{sec:decomp} focuses on the radiative transfer method we use to derive the physical properties of CNM and WNM. In Section~\ref{sec:result} we present our results on the properties of the CNM and WNM  (e.g. spin temperature, column density and CNM fraction) in the SMC/LMC outskirts. Section~\ref{sec:discuss} discusses our comparison of HI properties to other HI absorption surveys which focus mainly on the main body of the SMC/LMC, and a consideration of local environments for the cold gas in the SMC and the LMC. Section~\ref{sec:conslu} summarizes our main conclusions.

\section{Observations} \label{sec:obs}
The Magellanic Clouds observations were conducted by the GASKAP-HI survey. The SMC was part of the GASKAP Pilot Phase I observations with two 12-hour observations per field (ASKAP scheduling blocks 10941 and 10944) in December 2019, using the standard GASKAP-HI observing configuration \citep[see][]{Pingel2022}. The field was centered on J2000 RA = 00h58m43.280s, Dec = $-$72d31m49.02s. The spectral resolution is $ \sim 0.24~ \text{km}~ \text{s}^{–1}$. The observed band covered 18.5 MHz centered on 1419.85 MHz with 15558 channels, however, only the 2048 channels covering the Milky Way and SMC LSR velocity ranges were processed. Our analysis of HI spectra will focus on the SMC velocity range from 60 $ \text{km}~ \text{s}^{–1}$ to 250 $ \text{km}~ \text{s}^{–1}$. 
For each observation, data were processed using the ASKAPSoft package \citep{Hotan2021} and WSClean software package \citep{Offringa2014}, which calibrated the data and generated a continuum image and a continuum source catalog. Detailed descriptions of these observations and the initial data processing are provided by \cite{Pingel2022}. 

The LMC data were collected as part of the GASKAP-HI Pilot Phase II survey, which targeted nine fields (five towards the LMC, three toward the Bridge, and one towards the SMC) from June 2020 to March 2022. Each field received 10 hours of observation using the standard GASKAP-HI configuration \citep[see][]{Pingel2022}, centering on its corresponding J2000 RA \& Dec coordinates. The spectral resolution is $ \sim 0.24~ \text{km}~ \text{s}^{–1}$. 
Our analysis of HI spectra of the LMC will focus on the LSR velocity range from 170 $ \text{km}~ \text{s}^{–1}$ to 325 $ \text{km}~ \text{s}^{–1}$. Utilizing the same data processing techniques as those applied to the SMC observations, these data were also calibrated and reduced via ASKAPsoft and were imaged using WSClean. 

As described in \cite{Pingel2022}, the GASKAP data cube for both the SMC and the LMC was later combined with single-dish HI observations from the Parkes Galactic All-Sky Survey \citep[GASS;][]{McClure-Griffiths2009}, giving a composite dataset with $30''$ angular resolution, velocity resolution of $\sim 1 ~ \text{km}~ \text{s}^{–1}$, and sensitivity of 1.1 K per channel.

\subsection{HI absorption spectra}
The absorption pipeline was developed by \cite{Dempsey2022} and was applied to both the GASKAP-HI Pilot I and the GASKAP-HI Pilot II surveys. Please see this work for specific details about the absorption pipeline. In brief, from the continuum source catalog generated by ASKAPsoft, sources with a flux density of $S_{cont}\geq 15 ~\text{mJy}$ were selected. Then for each source, a small ($50''\times50''$) spectral line cube was generated centered on the source. During the imaging process of this sub-cube, a 1.5 k$\lambda$ (315m) baseline length cutoff was used to achieve high signal-to-noise ratios and filter out as much extended emission (extended WNM) as possible, since our goal is to detect the HI absorption arising from the CNM. All baselines are included in the construction of the associated HI emission spectra to ensure accurate estimates of e.g., the spin temperature. Within this cube, on-source emission spectra were extracted for all pixels within the source ellipse defined by the source-finding algorithm, {\tt Aegean} \citep{Hancock2012, Hancock2018}. Next, a line-free region was defined in the data cube, and the mean continuum flux density was calculated for each pixel. As described in \cite{Dickey1992}, a weighted-mean on-source spectrum was created, with each pixel being weighted by the square of the pixel mean continuum flux density. Finally, the absorption spectrum ($e^{-\tau}$) was produced by dividing the combined emission spectrum by its mean continuum flux density (measured within a line-free region). 
These absorption data, initially collected at a spectral resolution of 0.24 $ \text{km}~ \text{s}^{–1}$, were  smoothed to achieve a spectral resolution of $\sim 1$ $ \text{km}~ \text{s}^{–1}$.

The noise in the absorption spectrum was measured by combining the noise in the off-line region and emission in the primary beam of the dish \citep{Jameson2019}. The base noise of the spectrum was given by the standard deviation of the spectrum in the line-free region. To model the increase in system temperature caused by emission received by the antenna at different frequencies, the emission data from the Parkes Galactic All-Sky Survey (GASS) were used. The GASS emission was averaged across a 7 pixel (33 arcmin) radius annulus centered on the source position, excluding the central pixel. The $1\sigma$ noise profile for the GASKAP absorption spectrum was estimated by:
\begin{equation}
\sigma_\tau(v)=\sigma_{\mathrm{cont}} \frac{T_{\mathrm{sys}}+\eta_{\mathrm{ant}} T_{\mathrm{em}}(v)}{T_{\mathrm{sys}}}
\end{equation}
where $T_{\mathrm{sys}}=50$ K is the system temperature and $\eta_{\mathrm{ant}}=0.67$ is the antenna efficiency \citep{Hotan2021}, $\sigma_{\mathrm{cont}}$ is the standard deviation of the line-free region of the spectrum, and $T_{\mathrm{em}}(v)$ is the mean brightness temperature as measured in GASS.

Absorption features were identified in the spectra by one channel of $3\sigma$ absorption and an adjacent channel of $\geq 2.8\sigma$. The quality of identified features was also quantified by three tests: whether 1$\sigma$ noise level $<$ 1/3, whether the ratio of the deepest absorption to the highest emission noise $\geq 3$ and whether the range from the deepest absorption to highest emission noise $< 1.5$ \citep[details in][]{Dempsey2022}, classifying from A to D. Rating A spectra passed all tests, while the rating was reduced by one step for each failed test until rating D spectra failed all tests. 

\cite{Dempsey2022} provided absorption spectra towards 229 continuum sources in the direction of the SMC, with 65 HI absorption detections. Similarly, Dempsey et al. (in preparation) identified 1637 continuum sources with 222 HI absorption detections for the LMC. 

In the SMC, 85\% HI absorption detected sources (55 out of 65)  and 83\% sources (185 out of 222) in the LMC have a semi-major axis $<$ 7.5 arcsec  \citep[$\sim$ half beam size of the HI absorption datacube in][]{Dempsey2022}. A few sources in the LMC have semi-major axes exceeding 50 arcsec. Smaller source size than the beam size typically suggests a background extragalactic quasar, while a larger size may indicate a star-forming region within the galaxy.

\section{Target source selection}\label{sec:source_sel}

\begin{figure*}
    \centering
    \includegraphics[width=\textwidth]{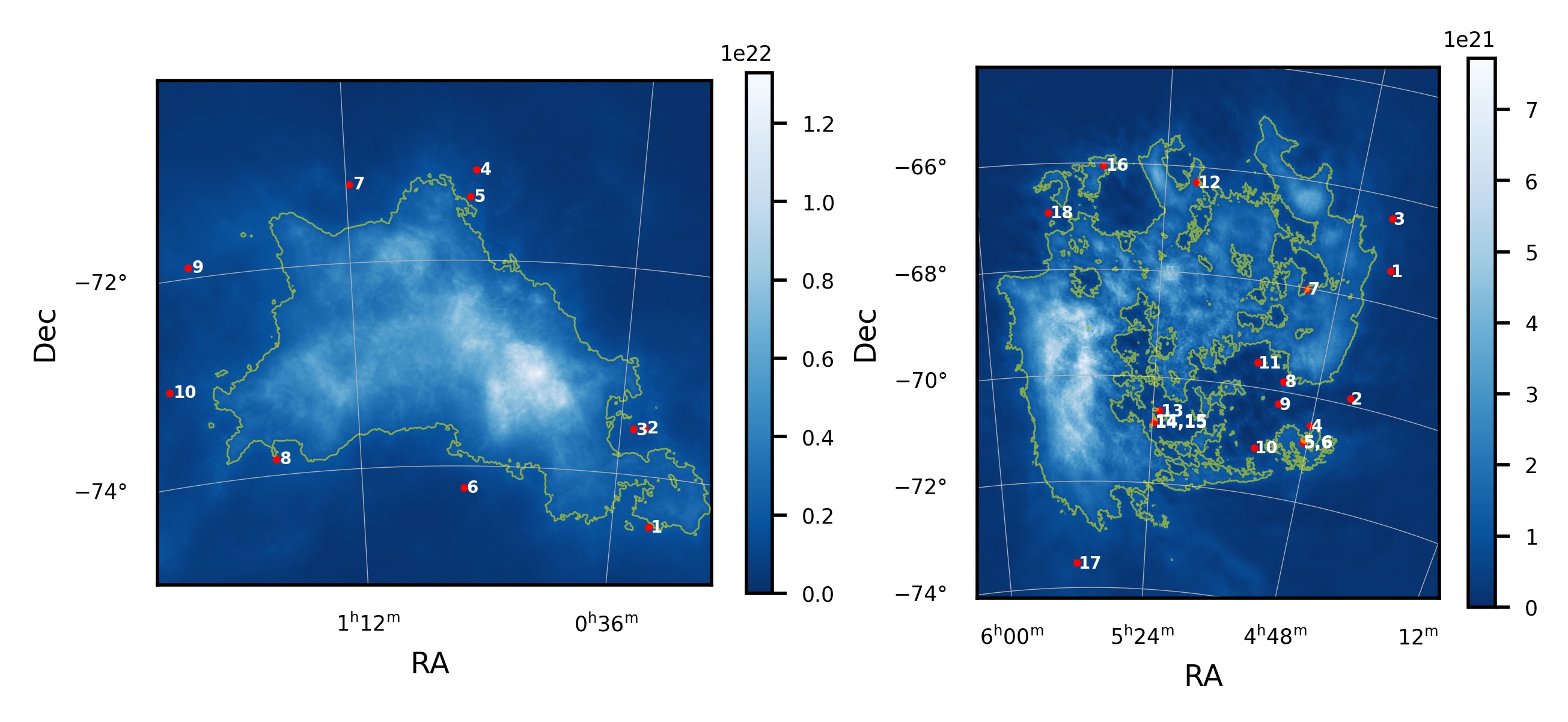}
    \caption{Selected background radio continuum sources for the SMC (left) and the LMC (right). The HI column density images are from the GASKAP phase I pilot survey \citep{Pingel2022} and  GASKAP phase II pilot survey with contours indicating $2\times10^{21} ~\text{cm}^{-2}$ for the SMC and $1\times10^{21} ~\text{cm}^{-2}$ for the LMC. Our target sources are colored as red numbered points with labels in the right. Note that some sources may overlap. These emission data are taken from archived data obtained through the CSIRO ASKAP Science Data Archive, CASDA (\url{https://research.csiro.au/casda}).}
    \label{fig:Map_source}
\end{figure*}

To select background continuum sources that probe outer, diffuse regions of the Clouds, we calculate their surrounding HI column density under the optically thin assumption \citep[see eq.3 in][]{Dickey1990}. 
For the SMC, we apply a column density threshold of $2\times10^{21} ~\text{cm}^{-2}$ and select sources in regions with column density lower than this threshold. While this is an arbitrary threshold, it was frequently used to separate the HI emission in the main body of the SMC from the more diffuse outskirts \citep[e.g.][]{Jameson2019,Pingel2022,Dempsey2022}. For the LMC, where the peak column density is lower than in the SMC, we adopt $1\times10^{21} ~\text{cm}^{-2}$ as the column density threshold \citep[e.g.][]{Liu2021}. 
We compare the HI column density in the direction of each source with the average HI column density within one beam size surrounding each source.  
We exclude sources with exceptionally faint HI absorption, specifically those where very few channels exceed the 3$\sigma$ threshold. For sources located very close to one another -- less than 15 arcsec apart, equivalent to the beam size of the HI absorption datacube -- we select only the one with the highest signal-to-noise ratio to avoid duplicate analysis due to correlated HI absorption spectra.
We finally select 10 background radio continuum sources from \cite{Dempsey2022} for the SMC and 18 sources from the LMC (Dempsey et al. in preparation).
Our samples contain about 20\% of the full samples provided for the SMC and the LMC.
All but one source (Source 12 in the LMC as shown in Figure~\ref{fig:Map_source} has a semi-major axis $\sim$ 77 arcsec) are background radio continuum sources and therefore probe the entire line-of-sight through the SMC/LMC.
 
We show selected sources in Figure~\ref{fig:Map_source}, overplotted on the HI column density images. Most of the selected sources in the SMC are located in the far outskirts of the SMC main body. However, for the LMC, while several sources are located in the diffuse outskirts of the LMC, many are within the main body of the LMC and probe the vicinity of large bubble-like structures.

We detect cold HI absorption at distances up to 3.4 kpc from the SMC's kinematic center and 4.3 kpc from the LMC's kinematic center. We used the kinematic centers of the SMC/LMC measured by \cite{Stanimirovic2004} and \cite{Kim1998}, respectively.

\subsection{HI emission spectra}\label{sec:HIemi}
To calculate the physical properties of the CNM and WNM, we derive the corresponding HI emission spectrum in the direction of each selected continuum source.

The HI emission data for both the SMC and LMC are obtained from the GASKAP-HI survey. 
For radiative transfer calculations, we require an HI emission profile that would be observed in the direction of background sources if the continuum sources were absent, the so called “expected” HI emission profile based on \cite{Heiles2003}.
We therefore calculate the HI emission spectrum for each source as the mean spectrum within an annulus centered on the source position, with an outer radius of $56''$ (8 pixels, $\sim 2 $ beam widths) and an inner exclusion radius of $28''$ (4 pixels, $\sim 1 $ beam width). We derive the 1$\sigma$ noise level from the standard deviation of emission spectra within the annulus.
Specifically, for Source 6 in the SMC, we adjust the size of the annulus to an outer radius of 10 pixels and an inner radius of 8 pixels due to the continuum source appearing extended and therefore affecting HI emission spectra.
We note that the radiative transfer calculations that we present in Section~\ref{sec:decomp} assume that the HI absorption and emission spectra sample the same HI structures.
While this assumption is often true in the Milky Way where in emission spatial resolution of sub-pc can be achieved, the ASKAP beamwidth of $30''$ corresponds to approximately 7--9 pc in the Magellanic Clouds due to their greater distance from us. Therefore, the emission derived here may not trace the exact same regions as the absorption. We will further discuss this effect on radiative transfer calculations in Section~\ref{sec:beam}.

\section{Gaussian Decomposition and Radiative Transfer Calculations}\label{sec:decomp}

After selecting the radio continuum sources and obtaining the corresponding emission and absorption spectra, we decompose the spectra to investigate the physical properties of the individual clouds along the line-of-sight. The method we use here originates from \cite{Heiles2003} and assumes that HI absorption spectrum (1-$e^{-\tau}$) mainly originates from the CNM, while HI emission spectrum (brightness temperature) is a product of both CNM and the WNM.
The total expected brightness temperature can be expressed as:
\begin{equation}
T_{\exp }'(v)=T_{B, \mathrm{CNM}}(v)+T_{B, \mathrm{WNM}}(v),
\end{equation}

We start by fitting the optical depth spectrum with a set of $N$ Gaussian functions, corresponding to $N$ CNM components:
\begin{equation}\label{eq:tau}
\tau(v)=\sum_0^{N-1} \tau_{0, n} e^{-4 \ln 2 \left[\left(v-v_{0, n}\right) / \delta v_n\right]^2},
\end{equation}
where $N$ is the minimum number of components necessary to make the residuals of the fit smaller or comparable to the estimated noise level of $\tau(v)$. $\tau_{0, n}$ is the peak optical depth, $v_{0, n}$ is the central velocity, and $\delta v_n$ is the FWHM of component $n$ . 
Then, the corresponding emission from $N$ CNM components can be represented as:
\begin{equation}
T_{B, \mathrm{CNM}}(v)=\sum_0^{N-1} T_{s, n}\left(1-e^{-\tau_n(v)}\right) e^{-\sum_0^{M_n-1} \tau_m(v)},
\end{equation}
where $\tau_m(v)$ represents each of the $M_n$ CNM clouds that lie in front of cloud $n$ and we will consider all possible orders of the CNM clouds. $T_{s,n}$ represents the derived spin temperature for each order.

For the WNM part in the emission profile, we use $K$ Gaussians to represent the original unabsorbed emission from WNM. To account for the absorption by CNM, we assume that a fraction $(1-F_k)$ of WNM is located behind all CNM components and the rest remains unabsorbed. The WNM brightness temperature is given by:
\begin{align}
T_{B, \mathrm{WNM}}(v) &= \sum_0^{K-1}\bigg[F_k+\left(1-F_k\right) e^{-\tau(v)}\bigg] \times \nonumber \\
&\quad T_{0, k} e^{-4 \ln 2 \left[\left(v-v_{0, k}\right) / \delta v_k\right]^2},
\end{align}
where $T_{0, k}$, $v_{0, k}$ and $\delta v_k$ represent the Gaussian fitting parameters (peak in units of brightness temperature, central velocity, FWHM) of original unabsorbed $k$-th WNM. We examine three specific values for the fraction $F_k: (0, 0.5, 1)$. The two extremes where $F_k=0$ and $F_k=1$ correspond to all WNM lying behind the CNM with total absorbed WNM and all WNM being in front of CNM with unabsorbed WNM, respectively. We set the minimum $T_{0, k}$ to be the average of the 1-$\sigma$ noise level of the emission spectra.

It is important to note that diffuse radio continuum emission (referred to as $T_{\text{sky}}$), including the CMB and the synchrotron emission, is always present but typically weak. Being much weaker than the Milky Way HI emission, this contribution can sometimes be ignored in the Milky Way studies \citep[e.g.][]{Heiles2003,Murray2015,Murray2018}. However, our analysis focuses on the outskirts of the LMC/SMC, where the brightness temperature is sometimes exceptionally low, less than 10 K. In these regions, the influence of $T_{\text{sky}}$ on the total brightness temperature is non-negligible. To account for this, we add an additional term into the expected brightness temperature, representing the blank sky background temperature attenuated by the absorbing HI cloud. Then, $T_{\text {sky }}$ should be removed considering baseline subtraction during the data reduction process. Therefore, our final total expected brightness temperature $T_{exp}(v)$ is expressed as:
\begin{equation}\label{eq:Texp}
    T_{exp}(v)=T_{exp}'(v)+T_{\text {sky }}e^{-\tau(v)}-T_{\text {sky }},
\end{equation}
where $e^{-\tau}$ represents the attenuation by CNM absorption and the last term subtracts $T_{\text{sky}}$ for the baseline correction. Same considerations were employed in \cite{Lee2015}.
The synchrotron emission maps of the Magellanic Clouds from \cite{Hassani2022} indicate a very low level of synchrotron emission in our selected directions (often $< 0.1 K$). Thus, here we only include the CMB contribution and set $T_{\text{sky}}=T_{\text{CMB}}=2.73$ K \citep{Fixsen2009}. 


\subsection{Determining the best fit}
The fitting process begins by modeling the optical depth using $N$ Gaussian components as described in Equation~\ref{eq:tau}. With these $N$ CNM components, we proceed to fit Equation~\ref{eq:Texp} to determine the spin temperature of each CNM components and the number and properties of WNM components.

The best fit in the Gaussian fitting process involves many considerations,  one of which is the number of fit components. 
To determine the optimal number of Gaussian functions needed for the best fit and avoid overfitting, we apply the Bayesian information criterion (BIC). The BIC quantifies the likelihood of the set of parameter values found given the goodness of the spectral fit and includes a penalty term that increases as the number of parameters used increases. For further insights into the BIC and other informational criteria, refer to \cite{Liddle2007}. 
The BIC is defined as follows:
\begin{equation}
    \text{BIC}=p \ln (N_d)-2 \ln (\hat{L}),
\end{equation}
where $p$ is the number of parameters in the model, $N_d$ is the number of data points, $\hat{L}$ is the maximized value of the likelihood function of the model.
We can approximate $\hat{L}$ for normally distributed errors as:
\begin{equation}
    \hat{L} \approx \exp \left(-\chi^2 / 2\right),
\end{equation}
where $\chi^2$ is defined as:
\begin{equation}
\chi^2=\sum\left(\frac{T_{exp}-T_{model }}{\sigma_{exp}}\right)^2,
\end{equation}
with $T_{exp}$ representing the observed data, $T_{model}$ the fitted model data, and $\sigma_{exp}$ the noise in the emission spectrum. Substituting this into the BIC expression gives,
\begin{equation}
    \text{BIC}=p \ln (N_d)+\chi^2.
\end{equation}

The optical depth fitting process begins with a single Gaussian component, and the BIC is calculated. At each iteration, an additional Gaussian is added, and the BIC is recalculated. The process continues until the BIC reaches its minimum, at which point the corresponding number of Gaussian components (i.e., $N$ CNM components) is selected.

For the subsequent fit in Equation~\ref{eq:Texp}, we begin with zero WNM components, then incrementally add one more at each iteration. 
For each iteration with $N$ CNM components and $K$ WNM components, there are a total of $N$! possible orderings for the CNM and $3^k$ combinations for the WNM, regarding three possible values for $F_k$. Thus, the total number of possible combinations is $N!$ $3^k$. We calculate the BIC for each combination and select the minimum BIC as the representative BIC for this iteration.
This iteration process ceases when the representative BIC no longer shows improvement (or reduction), at which point the previous iteration's number of Gaussian components is selected as the best operation and the model with the least BIC is chosen as our best fit. 

For each fitting model, we set the minimum spin temperature for each CNM component at 2.73 K, equivalent to $T_{\text{sky}}$. For the clouds with very low density, in the absence of collisions, the spin temperature would converge to $T_{\text{sky}}$ \citep{Draine2011}.
The final spin temperatures $T_s$ and the corresponding errors for each CNM component are calculated by a weighted average over all  $N!$ $3^k$ trials of $N$ absorption components and $K$ WNM components that yield the least BIC (cf. Equations (21a) and (21b) of \cite{Heiles2003}).

During the fitting process, we allow a velocity shift of up to $\pm$ 4 $ \text{km}~ \text{s}^{–1}$ (that is, 4 channels) between the CNM components in the emission and absorption spectra to account for small velocity fluctuations. Besides turbulent velocity fluctuations, beam dilution can also contribute to small velocity offsets between emission and absorption components as the GASKAP beam size spans about 7--9 pc  at the distance of the Magellanic Clouds. In fact, our emission spectra were derived from an area $\sim$ 14--18 pc in size, and can exhibit velocity gradients relative to the region where absorption originates. These velocity gradients can arise from interstellar turbulence, but also systematic motions from gravitational  collapse, and/or stellar activity.
Thus, we allow a slightly broader range of velocity shifts when we fit the CNM components identified in the absorption spectra to the emission spectra in the LMC/SMC compared to the shifts commonly used for  Milky Way absorption studies \citep[often just one channel, e.g.][]{Murray2018, Nguyen2024}.

\subsection{Fitting results}\label{sec:result}

\begin{figure}
    \centering
    \includegraphics[width=0.5\textwidth]{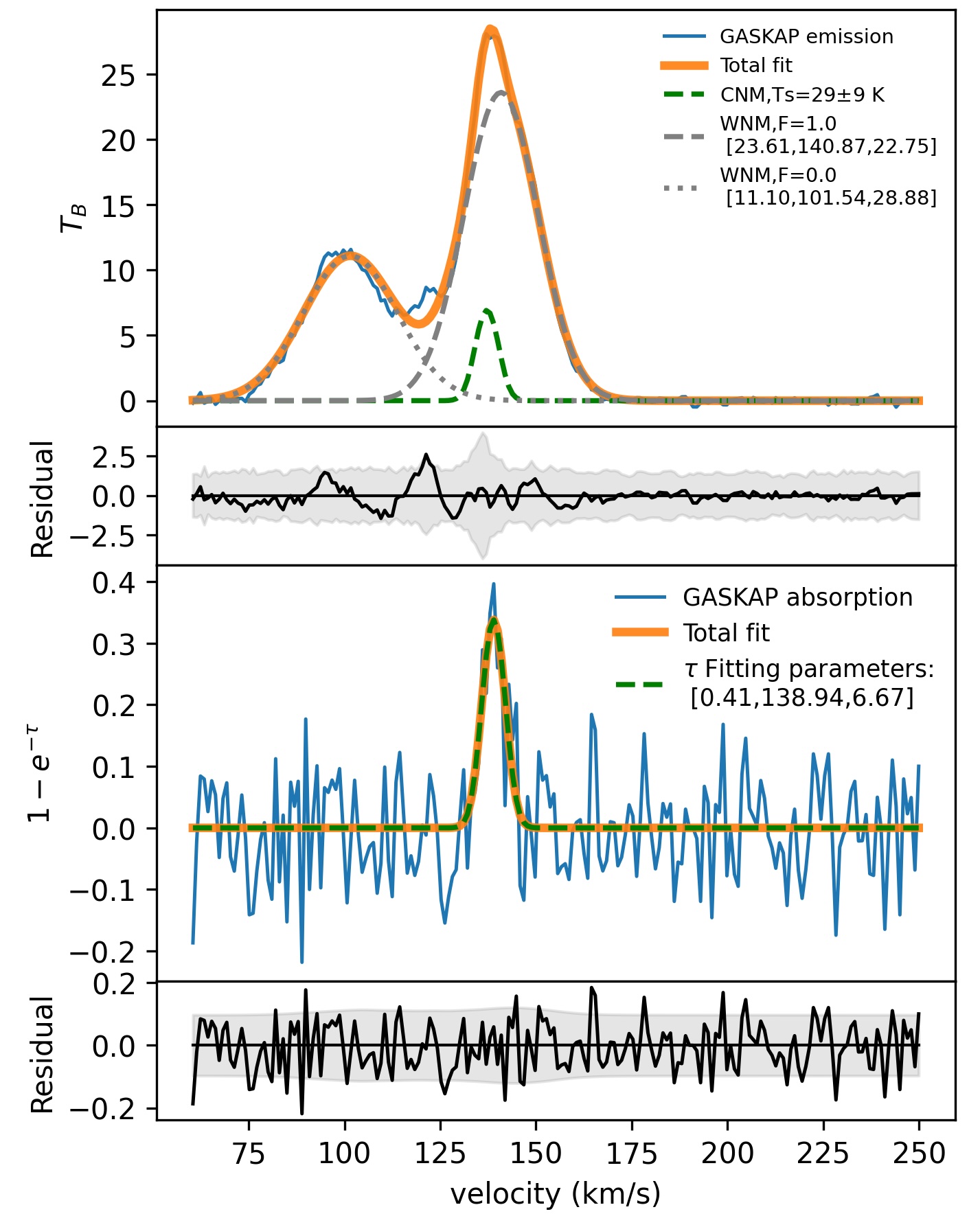}
    \caption{Gaussian decomposition for Source 1 from the SMC. The top panel shows the HI emission spectrum and its fitting residuals with shaded area presenting $\sigma_{exp}$ noise of the emission spectrum. The bottom panel presents the HI optical depth (absorption spectrum) and its fitting residuals with shaded area presenting $\sigma_{\tau}$ noise of the absorption spectrum. The total fits for both the emission and absorption spectra are shown in orange, while the individual CNM components are represented in green, and the individual WNM components are shown in gray. Following the 'CNM' label, the weighted mean spin temperature is presented. The 'F' value following each 'WNM' label quantifies the fraction of WNM in front of CNM, as stated in the text. The three values in every square brackets are, respectively, the amplitude, the central velocity, and the FWHM of each fitted Gaussian component.  All spectral fits are presented in Appendix~\ref{sec:all_spec}.}
    \label{fig:spectra_single_SMC}
\end{figure}

In Figure~\ref{fig:spectra_single_SMC}, we show an example of HI absorption and emission spectra as well as the corresponding Gaussian decomposition into CNM and WNM components. All fitting results are shown in Figures~\ref{fig:spectra_SMC}--\ref{fig:spectra_LMC2} (see Appendix~\ref{sec:all_spec}). For the majority of sources, the HI absorption spectrum requires only one or two Gaussian functions to produce a good fit. For the HI emission spectrum, we typically require 2--3 Gaussian components.   

The comparisons between emission and absorption spectra show that the SMC's emissions are complex and broad, while absorption detections have narrow linewidth and usually only one spectral feature. This is partially caused by the beam dilution and the fact that HI emission covers regions that are 14--18 pc in size, while HI absorption originates from a narrow solid angle traced by the background source. Another potential reason is 
a long line-of-sight depth of the SMC \citep{ Stanimirovic2004, deGrijs2015,Murray2024}, where the emission components may be physically separated along the line-of-sight, with only some of them harboring the CNM.
In the LMC, though the majority of the emission spectra are broad, there are several sources that display narrow emission spectra closely matching the corresponding absorption spectra (e.g. LMC Sources 2, 9). This suggests that in some locations in the LMC outskirts, the CNM structures can have a significant spatial extent.

We calculate the column density for each CNM and WNM component. For the CNM we use:
\begin{equation}
N_{\text{HI, CNM, }comp}=1.823\times 10^{18} \ T_s \int \tau(v) \ \mathrm{d} v \quad\mathrm{cm}^{-2},
\end{equation}
where $T_s$ and $\tau(v)$ are the spin temperature and optical depth of each component. We assume that WNM components are optically thin and calculate the column density using:
\begin{equation}
N_{\text{HI, WNM, }comp}=1.823\times 10^{18} \int T_{B,\text{WNM}}(v) \ \mathrm{d} v \quad\mathrm{cm}^{-2},
\end{equation}
where $T_{B,\text{WNM}}(v)$ is the brightness temperature. The total CNM/WNM column density for each LOS is the sum of all CNM/WNM components, referred to as $N_{\text{HI, CNM, }all}$ and $N_{\text{HI, WNM, }all}$. 

We estimate the CNM fraction along each line-of-sight by:
\begin{equation}
    f_{\text{CNM}}=\frac{N_{\text{HI, CNM, }all}}{N_{\text{HI, WNM, }all}+N_{\text{HI, CNM, }all}}.
\end{equation}

We also calculate the maximum kinetic temperature for each component via:
\begin{equation}\label{eq:Tk}
T_{k, \max }=\frac{m_{\mathrm{H}}}{8 k_{\mathrm{B}} \ln 2} \delta v_0^2=21.866 \cdot \delta v_0^2,
\end{equation}
where  $m_{\mathrm{H}}$ is the hydrogen mass and  $k_{\mathrm{B}}$ is the Boltzmann constant \citep{Draine2011}.

In Table~\ref{tab:SMC} and Table~\ref{tab:LMC}, we list the properties of the CNM/WNM cloud for our selected sources in the SMC and LMC. For each source, we list the source identification number, source position (RA, Dec) and the CNM fraction for each line-of-sight. For each CNM component, these tables list the spin temperature, the order of this component (starting from 0), the peak optical depth, central velocity, FWHM,  and the HI column density. For each WNM component, we list the F value, the peak brightness temperature, central velocity, FWHM and the HI column density.

\subsection{Beam dilution}\label{sec:beam}

As mentioned in Section~\ref{sec:HIemi}, one beam size covers a much larger physical area in the SMC/LMC than in the Milky Way. The emission spectra derived for each selected region include contributions from an area that is up to two beam sizes large. Therefore, due to the complex cloud structure, the emission in the LMC/SMC may not trace the same regions responsible for the HI absorption. This mismatch can lead to emission spectra that either miss or underestimate certain CNM components identified in absorption, or even display CNM components that do not appear in absorption. This issue can be alleviated somewhat with future higher-resolution observations, however, realistic resolution of HI emission is always lower than the resolution traced by pencil-beam HI absorption.

A key assumption in radiative transfer calculations (Section~\ref{sec:decomp}) is that the emission and absorption spectra trace the same interstellar structure. However, in the above cases, the emission spectra can sometimes fail to capture the features identified in the absorption spectra, which affects the accuracy of the derived parameters in radiative transfer calculations. 

This mismatch between emission and absorption spectra has also been seen in the Milky Way studies \citep[e.g.][]{Murray2018} and has resulted in unrealistically low estimates of the spin temperature.
The same effect occurs in this study, and components significantly impacted tend to have very low spin temperature, often $<10$ K. 
For example, Source 5 and Source 6 in the LMC probe regions within $\sim$ 30'' of each other. Both emission and absorption spectra for these sources are very similar. However, while absorption spectra have two sharp components (peaks), the emission spectra show only one corresponding strong component. As a result, this unmatched component has $T_s\sim5$ K, which is very low.
Additionally, Source 6 in the SMC, as well as Source 3 in the LMC, are likely affected by the emission-absorption beam mismatch, exhibiting low resolution emission spectra and similarly low $T_s$. 
4/34 CNM components in Tables 1 and 2 show $T_s < 10$ K. They are likely affected by beam dilution and their estimated spin temperatures are lower limits. We do not exclude these components from our analysis as we mostly focus on global sample comparisons.

{
\begin{longtable*}{C{0.5cm} C{1.2cm} C{0.7cm} C{1.1cm} C{1cm} C{1.1cm} C{0.6cm} C{1.8cm} C{1.8cm} C{2cm} C{0.8cm} C{2cm}}

\caption{SMC sources}\label{tab:SMC}\\
\hline
Label & Source & Ra & Dec &$f_{\text{CNM}}$ & $T_s$ & O & $\tau_{\text{peak}}$ or $T_B$ & $v_{\text{peak}}$ & FWHM & $T_{k,max}$ & $N_{\text{HI}}$ \\
& name & (deg) & (deg) & \% & (K) & or F&  or (K) & ($ \text{km}~ \text{s}^{–1}$) & ($ \text{km}~ \text{s}^{–1}$)& (K) & $10^{20}\text{cm}^{-2}$ \\ \hline
\endfirsthead

\multicolumn{12}{c}%
{{\bfseries Table \thetable\ continued from previous page}} \\
\hline
Label & Source & Ra & Dec &$f_{\text{CNM}}$ & $T_s$ & O & $\tau_{\text{peak}}$ or $T_B$ & $v_{\text{peak}}$ & FWHM & $T_{k,max}$ & $N_{\text{HI}}$ \\
& name & (deg) & (deg) &\% & (K) & or F&  or (K) & ($ \text{km}~ \text{s}^{–1}$) & ($ \text{km}~ \text{s}^{–1}$)& (K) & $10^{20}\text{cm}^{-2}$ \\ \hline
\endhead

\hline \multicolumn{12}{r}{{Continued on next page}} \\ \hline
\endfoot

\hline
\endlastfoot
1 &J003037-742903 & 7.654 & -74.484& 9$\pm$ 3 & 29 $\pm$ 9& 0 & 0.41 $\pm$ 0.00  & 138.9 $\pm$ 0.1  & 6.67 $\pm$ 0.28 & 974 & 1.55 $\pm$ 0.50 \\  
                & &  &  & & &1.0 & 23.61 $\pm$ 0.09 & 140.9 $\pm$ 0.0 & 22.75 $\pm$ 0.03 & 11319 & 10.42 $\pm$ 0.00 \\ 
                
                & &  &  & & &0.0 & 11.10 $\pm$ 0.03 & 101.5 $\pm$ 0.1 & 28.88 $\pm$ 0.15 & 18237 & 6.22 $\pm$ 0.00 \\ 
                \hline
        2 &J003242-733153 & 8.176 & -73.531& 6$\pm$ 3 & 19 $\pm$ 9& 0 & 1.16 $\pm$ 0.01  & 111.1 $\pm$ 0.0  & 2.24 $\pm$ 0.01 & 109 & 0.96 $\pm$ 0.44 \\  
                & &  &  & & &0.0 & 10.60 $\pm$ 0.16 & 146.7 $\pm$ 0.0 & 9.37 $\pm$ 0.07 & 1918 & 1.93 $\pm$ 0.00 \\ 
                
                & &  &  & & &0.5 & 11.22 $\pm$ 0.36 & 101.0 $\pm$ 0.3 & 28.36 $\pm$ 0.15 & 17585 & 6.17 $\pm$ 0.04 \\ 
                
                & &  &  & & &0.0 & 4.60 $\pm$ 0.05 & 157.0 $\pm$ 0.9 & 28.36 $\pm$ 0.75 & 17583 & 2.53 $\pm$ 0.01 \\ 
                
                & &  &  & & &0.0 & 18.98 $\pm$ 0.43 & 108.2 $\pm$ 0.0 & 13.08 $\pm$ 0.07 & 3742 & 4.82 $\pm$ 0.01 \\ 
                \hline
        3 &J003414-733329 & 8.559 & -73.558& 3$\pm$ 1 & 34 $\pm$ 9& 0 & 0.18 $\pm$ 0.00  & 108.0 $\pm$ 0.0  & 3.49 $\pm$ 0.09 & 266 & 0.41 $\pm$ 0.10 \\  
                & &  &  & & &1.0 & 12.21 $\pm$ 0.16 & 103.4 $\pm$ 0.1 & 34.03 $\pm$ 0.19 & 25328 & 8.07 $\pm$ 0.01 \\ 
                
                & &  &  & & &0.0 & 7.48 $\pm$ 0.04 & 160.4 $\pm$ 0.1 & 20.75 $\pm$ 0.17 & 9412 & 3.01 $\pm$ 0.00 \\ 
                
                & &  &  & & &1.0 & 14.55 $\pm$ 0.44 & 108.6 $\pm$ 0.0 & 10.04 $\pm$ 0.09 & 2204 & 2.83 $\pm$ 0.01 \\ 
                \hline
        4 &J005611-710707 & 14.047 & -71.119& 7$\pm$ 1 & 52 $\pm$ 10& 0 & 0.10 $\pm$ 0.00  & 141.5 $\pm$ 0.0  & 4.16 $\pm$ 0.03 & 378 & 0.41 $\pm$ 0.08 \\  
                & &  & & & 27 $\pm$ 10 & 1 & 0.09 $\pm$ 0.00  & 147.4 $\pm$ 0.0  & 3.78 $\pm$ 0.03 & 313 & 0.18 $\pm$ 0.07 \\

                & &  &  & & &0.0 & 22.91 $\pm$ 0.13 & 145.2 $\pm$ 0.0 & 17.82 $\pm$ 0.01 & 6943 & 7.92 $\pm$ 0.00 \\ 
                \hline
        5 &J005652-712300 & 14.219 & -71.384& 4$\pm$ 2 & 21 $\pm$ 13& 0 & 0.51 $\pm$ 0.00  & 148.6 $\pm$ 0.0  & 2.58 $\pm$ 0.02 & 145 & 0.52 $\pm$ 0.32 \\  
                & &  &  & & &0.5 & 27.02 $\pm$ 0.83 & 145.3 $\pm$ 0.0 & 15.42 $\pm$ 0.03 & 5198 & 8.09 $\pm$ 0.06 \\ 
                
                & &  &  & & &0.0 & 11.23 $\pm$ 0.87 & 149.2 $\pm$ 0.1 & 28.70 $\pm$ 0.20 & 18011 & 6.26 $\pm$ 0.24 \\ 
                \hline
        6 &J005732-741243 & 14.385 & -74.212& 3$\pm$ 1 & 7 $\pm$ 1 \tablenotemark{*}& 0 & 0.59 $\pm$ 0.00  & 139.7 $\pm$ 0.0  & 2.24 $\pm$ 0.00 & 109 & 0.19 $\pm$ 0.04 \\  
                & &  &  & & &0.0 & 6.47 $\pm$ 0.03 & 164.9 $\pm$ 0.0 & 15.81 $\pm$ 0.10 & 5463 & 1.98 $\pm$ 0.00 \\ 
                
                & &  &  & & &0.0 & 3.80 $\pm$ 0.02 & 151.8 $\pm$ 0.5 & 53.13 $\pm$ 0.58 & 61726 & 3.92 $\pm$ 0.00 \\ 
                \hline
        7 &J011134-711414 & 17.893 & -71.237& 5$\pm$ 0 & 15 $\pm$ 1& 0 & 0.36 $\pm$ 0.00  & 114.7 $\pm$ 0.1  & 4.00 $\pm$ 0.17 & 349 & 0.43 $\pm$ 0.04 \\  
                & &  &  & & &0.0 & 2.37 $\pm$ 0.07 & 117.1 $\pm$ 0.9 & 22.78 $\pm$ 3.12 & 11343 & 1.05 $\pm$ 0.02 \\ 
                
                & &  &  & & &1.0 & 9.52 $\pm$ 0.04 & 147.0 $\pm$ 0.0 & 15.55 $\pm$ 0.13 & 5289 & 2.87 $\pm$ 0.00 \\ 
                
                & &  &  & & &0.5 & 10.08 $\pm$ 0.03 & 175.0 $\pm$ 0.1 & 22.97 $\pm$ 0.15 & 11541 & 4.50 $\pm$ 0.00 \\ 
                \hline
        8 &J012349-735039 & 20.957 & -73.844& 7$\pm$ 1 & 33 $\pm$ 7& 0 & 0.41 $\pm$ 0.00  & 137.8 $\pm$ 0.0  & 4.71 $\pm$ 0.06 & 485 & 1.25 $\pm$ 0.26 \\  
                & &  &  & & &1.0 & 12.81 $\pm$ 0.02 & 170.3 $\pm$ 0.0 & 30.03 $\pm$ 0.07 & 19718 & 7.47 $\pm$ 0.00 \\ 
                
                & &  &  & & &0.0 & 19.21 $\pm$ 0.03 & 130.2 $\pm$ 0.0 & 21.73 $\pm$ 0.02 & 10327 & 8.10 $\pm$ 0.00 \\ 
                \hline
        9 &J013218-715348 & 23.078 & -71.897& 10$\pm$ 2 & 32 $\pm$ 7& 0 & 0.96 $\pm$ 0.01  & 213.7 $\pm$ 0.0  & 2.69 $\pm$ 0.03 & 158 & 1.62 $\pm$ 0.33 \\  
                & &  &  & & &0.0 & 15.20 $\pm$ 0.26 & 211.6 $\pm$ 0.0 & 9.18 $\pm$ 0.05 & 1843 & 2.71 $\pm$ 0.00 \\ 
                
                & &  &  & & &1.0 & 2.40 $\pm$ 0.01 & 133.0 $\pm$ 1.4 & 51.04 $\pm$ 3.68 & 56952 & 2.37 $\pm$ 0.03 \\ 
                
                & &  &  & & &1.0 & 17.46 $\pm$ 0.09 & 202.5 $\pm$ 0.1 & 27.28 $\pm$ 0.05 & 16268 & 9.24 $\pm$ 0.00 \\ 
                \hline
        10 &J013704-730413 & 24.269 & -73.070& 5$\pm$ 1 & 16 $\pm$ 3& 0 & 0.39 $\pm$ 0.00  & 190.3 $\pm$ 0.0  & 3.94 $\pm$ 0.05 & 339 & 0.47 $\pm$ 0.09 \\  
                & &  &  & & &1.0 & 8.33 $\pm$ 0.12 & 164.5 $\pm$ 1.4 & 30.44 $\pm$ 1.18 & 20258 & 4.92 $\pm$ 0.04 \\ 
                
                & &  &  & & &0.0 & 7.15 $\pm$ 0.20 & 192.0 $\pm$ 1.6 & 28.16 $\pm$ 1.27 & 17345 & 3.91 $\pm$ 0.04 \\ 
                \hline
        
\end{longtable*}
}
\tablenotetext{*}{$T_s< 10$ K CNM components, likely affected by beam dilution}

\begin{longtable*}{C{0.5cm} C{1.2cm} C{0.7cm} C{1.1cm} C{1cm} C{1.1cm} C{0.6cm} C{1.8cm} C{1.8cm} C{2cm} C{0.8cm} C{2cm}}
\caption{LMC sources}\label{tab:LMC}\\
\hline
 Label & Source & Ra & Dec &$f_{\text{CNM}}$ & $T_s$ & O & $\tau_{\text{peak}}$ or $T_B$ & $v_{\text{peak}}$ & FWHM & $T_{k,max}$ & $N_{\text{HI}}$ \\
& name & (deg) & (deg) & \% & (K) & or F&  or (K) & ($ \text{km}~ \text{s}^{–1}$) & ($ \text{km}~ \text{s}^{–1}$)& (K) & $10^{20}\text{cm}^{-2}$ \\ \hline
\endfirsthead

\multicolumn{12}{c}%
{{\bfseries Table \thetable\ continued from previous page}} \\
\hline
Label & Source & Ra & Dec &$f_{\text{CNM}}$ & $T_s$ & O & $\tau_{\text{peak}}$ or $T_B$ & $v_{\text{peak}}$ & FWHM & $T_{k,max}$ & $N_{\text{HI}}$ \\
& name & (deg) & (deg) &\% & (K) & or F&  or (K) & ($ \text{km}~ \text{s}^{–1}$) & ($ \text{km}~ \text{s}^{–1}$)& (K) & $10^{20}\text{cm}^{-2}$ \\ \hline
\endhead

\hline \multicolumn{12}{r}{{Continued on next page}} \\ \hline
\endfoot

\hline
\endlastfoot
1 &J043855-672153 & 69.733 & -67.365& 10$\pm$ 1 & 31 $\pm$ 4& 0 & 0.11 $\pm$ 0.00  & 241.0 $\pm$ 0.0  & 4.88 $\pm$ 0.03 & 520 & 0.32 $\pm$ 0.04 \\  
                & &  & & & 24 $\pm$ 4 & 1 & 0.08 $\pm$ 0.00  & 247.0 $\pm$ 0.0  & 3.71 $\pm$ 0.04 & 301 & 0.15 $\pm$ 0.03 \\

                & &  &  & & &1.0 & 9.34 $\pm$ 0.04 & 249.9 $\pm$ 0.0 & 23.75 $\pm$ 0.06 & 12336 & 4.30 $\pm$ 0.00 \\ 
                \hline
        2 &J044047-695217 & 70.199 & -69.872& 100$\pm$ 0 & 89 $\pm$ 4& 0 & 0.05 $\pm$ 0.00  & 239.5 $\pm$ 0.1  & 4.47 $\pm$ 0.24 & 436 & 0.41 $\pm$ 0.03 \\  \hline
        3 &J044056-662423 & 70.236 & -66.406& 2$\pm$ 0 & 3 $\pm$ 0 \tablenotemark{*}& 0 & 0.30 $\pm$ 0.00  & 228.9 $\pm$ 0.0  & 4.11 $\pm$ 0.00 & 369 & 0.07 $\pm$ 0.00 \\  
                & &  &  & & &0.5 & 3.57 $\pm$ 0.00 & 250.1 $\pm$ 0.1 & 40.68 $\pm$ 0.25 & 36190 & 2.82 $\pm$ 0.00 \\ 
                \hline
        4 &J044809-703144 & 72.041 & -70.529& 2$\pm$ 0 & 10 $\pm$ 2& 0 & 0.29 $\pm$ 0.00  & 232.3 $\pm$ 0.0  & 2.50 $\pm$ 0.09 & 136 & 0.14 $\pm$ 0.03 \\  
                & &  &  & & &1.0 & 3.95 $\pm$ 0.26 & 263.1 $\pm$ 0.3 & 18.13 $\pm$ 1.27 & 7187 & 1.39 $\pm$ 0.02 \\ 
                
                & &  &  & & &0.0 & 3.66 $\pm$ 0.39 & 251.4 $\pm$ 0.8 & 47.55 $\pm$ 2.42 & 49448 & 3.38 $\pm$ 0.16 \\ 
                
                & &  &  & & &0.5 & 8.88 $\pm$ 0.13 & 237.7 $\pm$ 0.0 & 11.10 $\pm$ 0.10 & 2693 & 1.91 $\pm$ 0.00 \\ 
                \hline
        5 &J044902-705155 & 72.259 & -70.865& 5$\pm$ 1 & 52 $\pm$ 8& 0 & 0.12 $\pm$ 0.00  & 234.6 $\pm$ 0.1  & 3.54 $\pm$ 0.14 & 273 & 0.42 $\pm$ 0.07 \\  
                & &  & & & 7 $\pm$ 5 \tablenotemark{*} & 1 & 0.11 $\pm$ 0.00  & 246.3 $\pm$ 0.1  & 4.60 $\pm$ 0.22 & 463 & 0.07 $\pm$ 0.05 \\

                & &  &  & & &0.0 & 23.17 $\pm$ 0.06 & 239.1 $\pm$ 0.0 & 15.26 $\pm$ 0.02 & 5090 & 6.86 $\pm$ 0.00 \\ 
                
                & &  &  & & &0.0 & 3.39 $\pm$ 0.04 & 248.5 $\pm$ 0.7 & 44.19 $\pm$ 0.95 & 42705 & 2.91 $\pm$ 0.01 \\ 
                \hline
        6 &J044903-705212 & 72.266 & -70.870& 6$\pm$ 1 & 65 $\pm$ 9& 0 & 0.11 $\pm$ 0.00  & 237.5 $\pm$ 0.0  & 4.57 $\pm$ 0.10 & 455 & 0.65 $\pm$ 0.09 \\  
                & &  & & & 5 $\pm$ 4 \tablenotemark{*} & 1 & 0.08 $\pm$ 0.00  & 246.0 $\pm$ 0.1  & 3.79 $\pm$ 0.17 & 313 & 0.03 $\pm$ 0.02 \\

                & &  &  & & &1.0 & 23.54 $\pm$ 0.12 & 239.0 $\pm$ 0.0 & 15.33 $\pm$ 0.02 & 5140 & 7.00 $\pm$ 0.00 \\ 
                
                & &  &  & & &0.0 & 3.30 $\pm$ 0.04 & 248.5 $\pm$ 0.7 & 43.69 $\pm$ 0.85 & 41740 & 2.80 $\pm$ 0.00 \\ 
                \hline
        7 &J045421-680056 & 73.590 & -68.016& 12$\pm$ 5 & 21 $\pm$ 9& 0 & 0.39 $\pm$ 0.00  & 260.1 $\pm$ 0.1  & 7.86 $\pm$ 0.13 & 1349 & 1.25 $\pm$ 0.53 \\  
                & &  &  & & &0.0 & 8.22 $\pm$ 0.25 & 257.2 $\pm$ 0.0 & 30.58 $\pm$ 0.24 & 20445 & 4.88 $\pm$ 0.02 \\ 
                
                & &  &  & & &0.5 & 18.54 $\pm$ 2.86 & 257.8 $\pm$ 0.3 & 11.69 $\pm$ 0.20 & 2989 & 4.21 $\pm$ 0.43 \\ 
                \hline
        8 &J045545-694830 & 73.941 & -69.808& 16$\pm$ 1 & 61 $\pm$ 4& 0 & 0.11 $\pm$ 0.00  & 262.3 $\pm$ 0.1  & 5.54 $\pm$ 0.17 & 670 & 0.71 $\pm$ 0.05 \\  
                & &  &  & & &0.0 & 3.79 $\pm$ 0.02 & 221.4 $\pm$ 0.1 & 17.95 $\pm$ 0.25 & 7041 & 1.32 $\pm$ 0.00 \\ 
                
                & &  &  & & &1.0 & 8.88 $\pm$ 0.08 & 258.6 $\pm$ 0.0 & 14.59 $\pm$ 0.06 & 4651 & 2.51 $\pm$ 0.00 \\ 
                \hline
        9 &J045608-701434 & 74.036 & -70.243& 75$\pm$ 2 & 18 $\pm$ 3& 1 & 3.55 $\pm$ 0.00  & 234.4 $\pm$ 0.0  & 2.04 $\pm$ 0.00 & 91 & 2.59 $\pm$ 0.47 \\  
                & &  & & & 24 $\pm$ 2 & 0 & 1.02 $\pm$ 0.00  & 232.8 $\pm$ 0.0  & 6.81 $\pm$ 0.01 & 1015 & 3.19 $\pm$ 0.30 \\

                & &  &  & & &1.0 & 5.81 $\pm$ 0.12 & 229.5 $\pm$ 0.1 & 17.22 $\pm$ 0.21 & 6482 & 1.94 $\pm$ 0.00 \\ 
                \hline
        10 &J045954-710737 & 74.979 & -71.127& 27$\pm$ 3 & 11 $\pm$ 2& 0 & 1.12 $\pm$ 0.00  & 239.5 $\pm$ 0.0  & 2.77 $\pm$ 0.01 & 168 & 0.66 $\pm$ 0.11 \\  
                & &  &  & & &0.5 & 4.27 $\pm$ 0.04 & 232.5 $\pm$ 0.2 & 21.24 $\pm$ 0.47 & 9866 & 1.76 $\pm$ 0.00 \\ 
                \hline
        11 &J050201-693151 & 75.507 & -69.531& 16$\pm$ 1 & 45 $\pm$ 3& 0 & 0.18 $\pm$ 0.00  & 243.5 $\pm$ 0.0  & 5.34 $\pm$ 0.01 & 622 & 0.87 $\pm$ 0.06 \\  
                & &  &  & & &1.0 & 8.83 $\pm$ 0.05 & 238.2 $\pm$ 0.0 & 14.55 $\pm$ 0.08 & 4629 & 2.49 $\pm$ 0.00 \\ 
                
                & &  &  & & &1.0 & 1.83 $\pm$ 0.02 & 243.7 $\pm$ 1.3 & 58.35 $\pm$ 4.62 & 74445 & 2.07 $\pm$ 0.03 \\ 
                \hline
        12 &J051819-661717 & 79.581 & -66.288& 17$\pm$ 4 & 21 $\pm$ 8& 0 & 1.44 $\pm$ 0.00  & 288.3 $\pm$ 0.0  & 3.26 $\pm$ 0.01 & 231 & 1.92 $\pm$ 0.70 \\  
                & &  & & & 17 $\pm$ 9 & 1 & 0.49 $\pm$ 0.00  & 294.0 $\pm$ 0.0  & 2.61 $\pm$ 0.04 & 148 & 0.42 $\pm$ 0.21 \\

                & &  &  & & &0.0 & 5.83 $\pm$ 0.01 & 250.0 $\pm$ 0.0 & 21.83 $\pm$ 0.10 & 10415 & 2.47 $\pm$ 0.00 \\ 
                
                & &  &  & & &0.0 & 6.23 $\pm$ 0.24 & 291.0 $\pm$ 0.1 & 32.69 $\pm$ 0.65 & 23366 & 3.95 $\pm$ 0.03 \\ 
                
                & &  &  & & &1.0 & 18.36 $\pm$ 0.19 & 289.4 $\pm$ 0.0 & 13.23 $\pm$ 0.04 & 3828 & 4.71 $\pm$ 0.00 \\ 
                \hline
        13 &J052229-703757 & 80.622 & -70.633& 17$\pm$ 1 & 54 $\pm$ 4& 0 & 0.27 $\pm$ 0.00  & 234.8 $\pm$ 0.0  & 4.58 $\pm$ 0.03 & 459 & 1.31 $\pm$ 0.11 \\  
                & &  &  & & &0.0 & 9.36 $\pm$ 0.02 & 236.0 $\pm$ 0.0 & 35.57 $\pm$ 0.11 & 27667 & 6.46 $\pm$ 0.00 \\ 
                \hline
        14 &J052340-705019 & 80.920 & -70.839& 7$\pm$ 1 & 47 $\pm$ 8& 0 & 0.23 $\pm$ 0.00  & 222.2 $\pm$ 0.0  & 2.99 $\pm$ 0.01 & 195 & 0.62 $\pm$ 0.11 \\  
                & &  &  & & &0.0 & 18.61 $\pm$ 0.07 & 224.8 $\pm$ 0.0 & 15.11 $\pm$ 0.03 & 4993 & 5.46 $\pm$ 0.00 \\ 
                
                & &  &  & & &0.0 & 2.26 $\pm$ 0.02 & 243.7 $\pm$ 3.2 & 57.03 $\pm$ 3.34 & 71123 & 2.51 $\pm$ 0.02 \\ 
                \hline
        15 &J052341-705122 & 80.921 & -70.856& 14$\pm$ 1 & 150 $\pm$ 7& 0 & 0.07 $\pm$ 0.00  & 227.9 $\pm$ 0.1  & 5.57 $\pm$ 0.14 & 678 & 1.15 $\pm$ 0.06 \\  
                & &  &  & & &0.0 & 15.22 $\pm$ 0.23 & 224.4 $\pm$ 0.0 & 17.48 $\pm$ 0.12 & 6681 & 5.16 $\pm$ 0.01 \\ 
                
                & &  &  & & &0.0 & 2.46 $\pm$ 0.02 & 246.5 $\pm$ 10.6 & 40.53 $\pm$ 10.51 & 35919 & 1.93 $\pm$ 0.25 \\ 
                \hline
        16 &J053544-660227 & 83.937 & -66.041& 6$\pm$ 5 & 21 $\pm$ 16& 0 & 0.33 $\pm$ 0.00  & 279.5 $\pm$ 0.0  & 5.11 $\pm$ 0.07 & 570 & 0.67 $\pm$ 0.54 \\  
                & &  &  & & &1.0 & 19.64 $\pm$ 0.09 & 290.5 $\pm$ 0.0 & 14.47 $\pm$ 0.04 & 4575 & 5.51 $\pm$ 0.00 \\ 
                
                & &  &  & & &1.0 & 6.09 $\pm$ 0.06 & 280.0 $\pm$ 0.3 & 39.49 $\pm$ 0.28 & 34101 & 4.66 $\pm$ 0.00 \\ 
                \hline
        17 &J054150-733215 & 85.461 & -73.538& 4$\pm$ 0 & 96 $\pm$ 4& 0 & 0.03 $\pm$ 0.00  & 253.8 $\pm$ 0.0  & 3.79 $\pm$ 0.03 & 313 & 0.21 $\pm$ 0.01 \\  
                & &  &  & & &0.0 & 6.27 $\pm$ 0.01 & 260.0 $\pm$ 0.1 & 34.77 $\pm$ 0.19 & 26439 & 4.23 $\pm$ 0.00 \\ 
                
                & &  &  & & &1.0 & 1.52 $\pm$ 0.01 & 199.5 $\pm$ 1.3 & 38.70 $\pm$ 4.10 & 32744 & 1.14 $\pm$ 0.01 \\ 
                \hline
        18 &J054623-665544 & 86.598 & -66.929& 37$\pm$ 5 & 36 $\pm$ 8& 0 & 0.54 $\pm$ 0.00  & 285.9 $\pm$ 0.2  & 7.66 $\pm$ 0.40 & 1284 & 2.95 $\pm$ 0.68 \\  
                & &  &  & & &1.0 & 15.21 $\pm$ 0.45 & 283.6 $\pm$ 0.0 & 16.74 $\pm$ 0.07 & 6129 & 4.94 $\pm$ 0.02 \\ 
                \hline
        
\end{longtable*}
\tablenotetext{*}{$T_s< 10$ K CNM components, likely affected by beam dilution}

\section{Physical properties of CNM/WNM HI in the outskirts of the SMC/LMC}
\begin{figure*}
    \centering
    \includegraphics[width=0.9\textwidth]{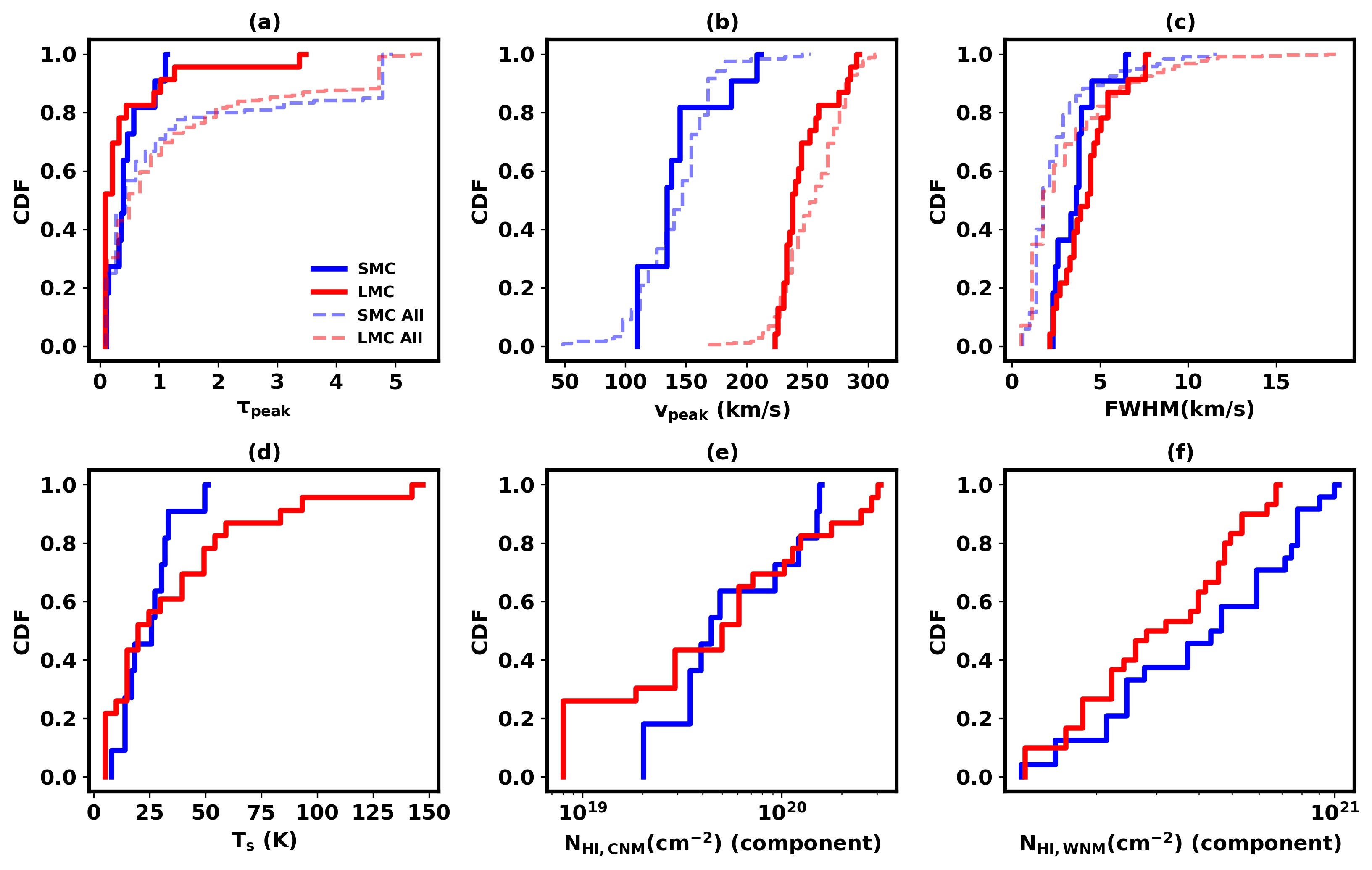}
    \caption{Cumulative distribution functions (CDF) of 
    fitted Gaussian components for selected SMC and LMC sources. Panels (a) through (e) show properties of the CNM components, while panel (f) shows properties of the WNM components. The top row, from left to right, shows the peak optical depth (a), peak velocity (b) and corresponding line width of CNM components (c). The lower panel, from left to right,  shows the spin temperature of the CNM (d), and the CNM HI column density (e) and the WNM HI column densities WNM (f)  for individual components. We also include the peak optical depth and peak velocity for all the HI absorption detected components in the SMC and LMC (dashed lines) as provided by \cite{Dempsey2022} and Dempsey et al. (in preparation), respectively. }
    \label{fig:property_CDF_com}
\end{figure*}

In this study, we analyzed 10 directions in the outskirts of the SMC, identifying 11 CNM components and 24 WNM components. In the LMC, from 18 directions, we identified 23 CNM components and 30 WNM components. 
Figure~\ref{fig:property_CDF_com} presents the cumulative distribution functions (CDFs) of the basic properties for all identified CNM (Panel (a) to (e)) and WNM components (Panel (f)). For comparison, we also include data products based on
the entire datasets of HI absorption samples from both the SMC and the LMC absorption catalogs \citep{Dempsey2022}.  

Panel (a) shows the distribution of the peak optical depth ($\tau_{\text{peak}}$) from the Gaussian fitting. The peak optical depth ranges from 0.09 to 1.16 with a median of 0.41 for the SMC, and 0.03 to 3.55 with a median of 0.23 for the LMC. 
Outskirts of both galaxies have lower $\tau_{\text{peak}}$ than the entire samples shown with dashed lines. The cold HI structures in the outskirts of the LMC have a broader range of $\tau_{\text{peak}}$ relative to what we find for the SMC outskirts. This result is not due to the sensitivity of the HI absorption spectra, as our median sensitivity in optical depth is similar for the SMC and LMC.  The median noise level for the optical depth is $\sigma_{\tau}=0.06$ with a range of $0.006\leq \sigma_\tau \leq 0.14$  for the SMC, and $\sigma_{\tau}=0.02$ with a range of $0.003\leq \sigma_\tau \leq 0.09$  for the LMC.
However, this is mostly caused by the selection bias where
all SMC spectra probe regions outside of the main body of the SMC, while some of the selected background sources probe regions in the LMC's main body. As shown in Figure~\ref{fig:property_map}, those directions tend to have higher optical depth. 
We also note that our median $\tau_{\text{peak}}$ of 0.4 
is in agreement with the pilot GASKAP-HI SMC studies by \citep{Dempsey2022} who found 16 HI absorption detections outside of the main body of the SMC with a median peak optical depth of 0.5.

Panel (b) presents the central velocity of the Gaussian absorption component. The peak velocities for our selected sources in the SMC align closely with those of the whole sample, while the LMC selected samples are shifted to the lower velocity range in comparison with the whole sample. This is most likely a selection bias, as most LMC samples are located in the south-west region where the global LMC rotation pattern has lower velocities (visible in the right column of Figure~\ref{fig:property_map}). 

Panel (c) shows the line widths (FWHM) of Gaussian absorption components. Our samples of SMC and LMC components have similar FWHM ranges. To compare these samples with the entire SMC/LMC absorption datasets we use the equivalent widths (EW) information provided by
\cite{Dempsey2022}. We assume that the line profiles of HI absorption features have Gaussian shape and convert the equivalent widths to FWHM via $\text{FWHM}=\text{EW}/(1.064\times \tau_{\text{peak}})$. 
Relative to the FWHM distributions derived for the entire samples, Gaussian CNM components in the outskirts of the SMC and LMC have typically higher line widths. This would be indicative of a higher kinetic temperature and/or turbulent broadening. 

Panel (d) illustrates the distribution of derived spin temperatures based on the Gaussian fitting for the SMC and the LMC. 
The spin temperatures mostly span a range from $\sim 20$ K to $\sim 50$ K (with a median of 27 K) for the SMC, and from $\sim10$ K to $\sim100$ K (with a median of 24 K) for the LMC.
The LMC sources display a wider range of spin temperatures compared to those in the SMC, suggesting a more complex and spatially varying environment in the LMC. This wide range of $T_s$ values in the LMC is likely influenced by the selection bias that several of the selected sources are situated within the LMC disk where they can be significantly impacted by internal stellar activities.
Comparing $T_s$ to the corresponding maximum kinetic temperature as shown in Table~\ref{tab:LMC} and Table~\ref{tab:SMC}, all the derived spin temperatures of the cold gas are significantly lower. We then calculate the sonic Mach number for each CNM component following equation (17) in \cite{Heiles2003b} and find an average of $\sim 7$ for both the SMC and the LMC, exceeding the Milky Way's Mach number of $\sim$ 3 \citep{Heiles2003b}.  There is no observed spatial dependence of the Mach number across our samples. 
While $T_s$ estimation is in the case of some of our sources  affected by the beam dilution, this result suggests highly turbulent motions of the CNM in both SMC and LMC.

Panel (e) and Panel (f) present the HI column density for each fitted CNM and WNM components, respectively. 
The CNM column densities of HI absorbing structures are comparable between the SMC and the LMC outskirts. However, the WNM column densities in the SMC are generally higher than in the LMC. These distributions are somewhat anticipated due to our selection criteria as well as SMC having a longer line-of-sight depth.
As a consequence, the CNM fraction is lower for the SMC outskirts, as shown in Figure~\ref{fig:property_CDF_LOS}.

\subsection{The CNM fraction}\label{sec:cnmF}
\begin{figure*}
    \centering
    \includegraphics[width=0.7\textwidth]{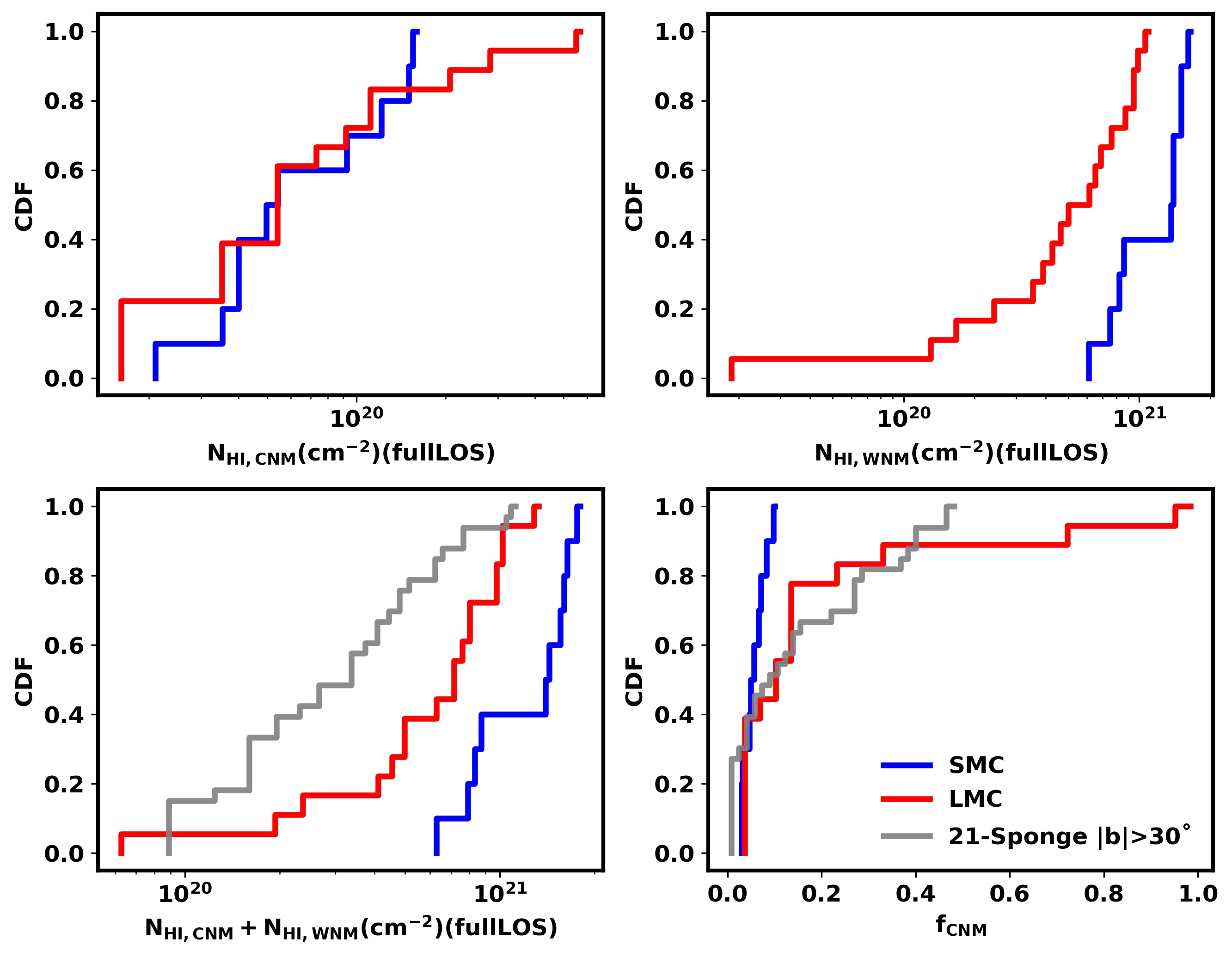}
    \caption{CDF of different properties of our selected sources in the LMC and SMC for all lines of sight. Upper left is the total CNM HI column density, upper right is the total WNM HI column density, lower left is the total HI column density and lower right is the CNM fraction.  For comparison, we include data from the 21-SPONGE survey of the Milky Way \citep{Murray2018}, selecting only lines of sight with Galactic latitude $|b|>30^\circ$ to represent diffuse regions in the Milky Way.}
    \label{fig:property_CDF_LOS}
\end{figure*}

\begin{figure}
    \centering
    \includegraphics[width=0.5\textwidth]{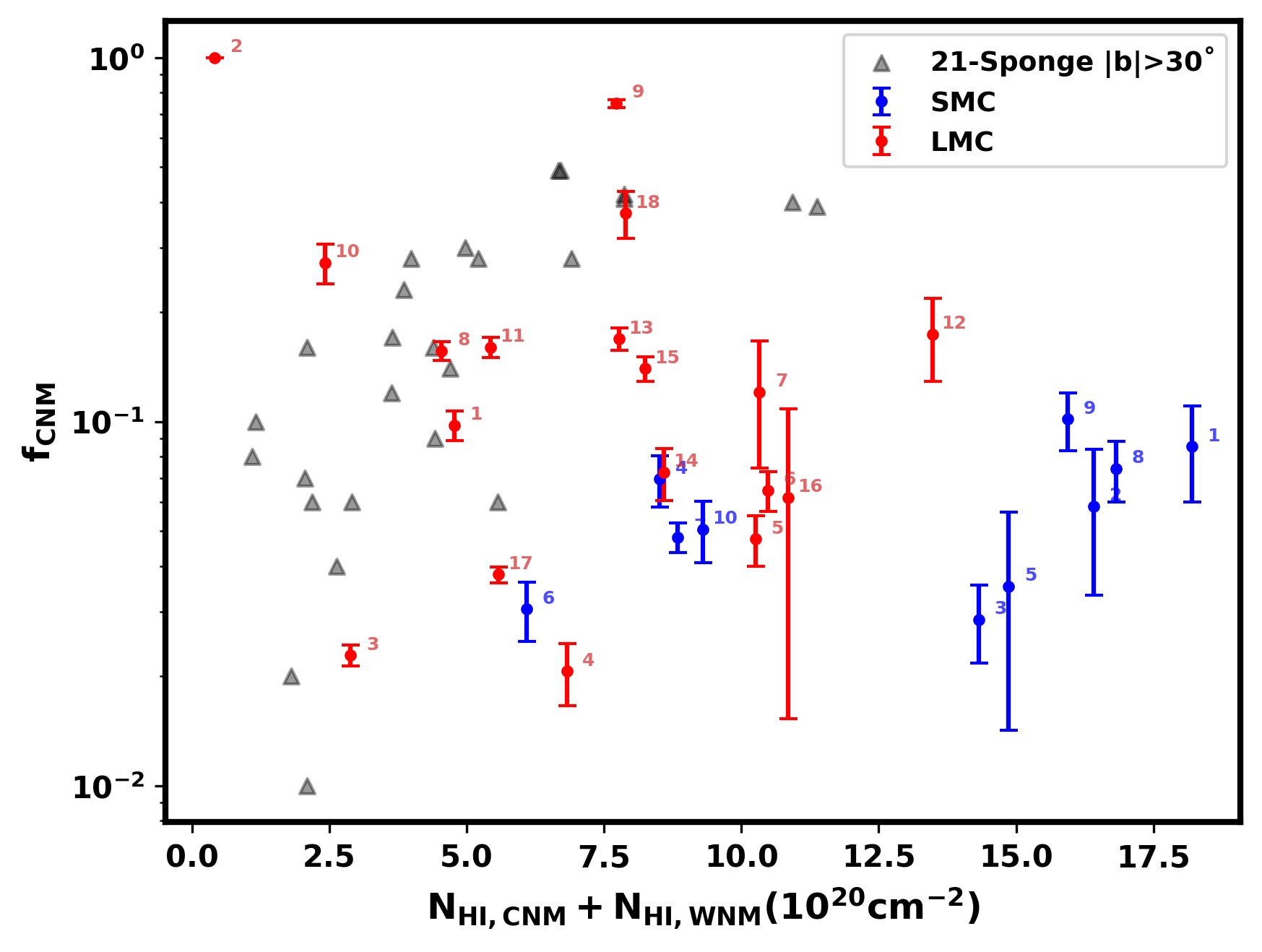}
    \caption{The CNM fraction as a function of the total HI column density. We also include data from the 21-SPONGE survey of the Milky Way \citep{Murray2018} with Galactic latitude $|b|>30^\circ$.}
    \label{fig:fCNM_NHI}
\end{figure}

For each line-of-sight, we calculate the total CNM and WNM HI column densities by summing up column density of individual components. Figure~\ref{fig:property_CDF_LOS} shows the column density distributions of the CNM, WNM and the CNM fraction. As already noticed, the general CNM HI column density of the LMC is comparable to that of the SMC, but the WNM and the total HI column density are smaller. This results in a higher fraction of cold HI in the LMC. The CNM fraction in the SMC has a very narrow distribution and ranges from 1\% to 11\% with a mean value of 6\%. In the LMC, the CNM fraction has a broad distribution and can reach up to 100\% with a mean value of 21\%. 

For comparison, we also include the HI absorption study of the Milky Way by \cite{Murray2018} (referred to as 21-SPONGE). We include 33 out of their 57 sources with
Galactic latitude $|b|>30^\circ$.
The 21-SPONGE generally has lower total HI column densities and shows a similar CNM fraction distribution as the LMC with $N_{\mathrm{HI}}<2\times 10^{21}$, though it only reaches up to 50\%.
Looking at an even larger HI absorption data set for the Milky Way from \cite{McClure-Griffiths2023}, the CNM fraction has been found to have a broad distribution but never gets higher than 80\%.  While the LMC's CNM fraction distribution is similar to that of the Milky Way, the LMC outskirts have directions where HI is 100\% in the form of CNM.

In Figure~\ref{fig:fCNM_NHI}, we show the CNM fraction as a function of the total HI column density for our selected sources. In the Milky Way, the CNM fraction typically increases with higher HI column densities \citep[see also Figure 8 of][]{Stanimirovic2014}. Similarly, our clouds in the SMC show a slight increase of the CNM fraction with increasing HI column density. However, the CNM fractions are significantly smaller than those in the LMC or Milky Way. The SMC has a much lower metallicity compared to the LMC and the Milky Way. In metal-poor environments, the ISM is expected to cool less efficiently due to the reduced abundance of key cooling agents such as CII and OI, and also has a more effective photoelectric heating \citep{Kim2024}. These effects naturally lead to a lower CNM fraction. Additionally, this difference could be due to a longer line-of-sight depth in the SMC \citep{ Stanimirovic2004, deGrijs2015,Murray2024}. As pointed out by \cite{Murray2024}, the SMC comprises two structures with distinct stellar and gaseous compositions along the line-of-sight. The SMC structure with a lower metallicity would be less abundant in the CNM.
Supporting this, in Figure~\ref{fig:spectra_SMC}, 80\% sightlines in the SMC (as opposed to 20\% in the LMC) contain WNM emission components without any CNM.

For the LMC, while the CNM fraction follows a similar trend to 21-SPONGE, the data points are more scattered. Notably, the cloud with the lowest HI column density ($\sim 0.5 \times 10^{20}$ cm$^{-2}$), labeled `2', shows the highest CNM fraction (equal to one). This cloud, located in the southeastern region of the LMC, has a nearby filamentary shell structure.  Additionally, cloud labeled `8', located at the edge of an expanding bubble in the LMC’s southwestern region, has a CNM fraction of $\sim$ 70\%, with the total HI column density of $\sim 8 \times 10^{20}$ cm$^{-2}$. These directions with a very high CNM fraction stand out in contrast to the SMC directions where the CNM fraction is $<11$\%.
In our selected regions, the CNM clouds in the LMC are often located near dynamic bubble-like structures driven by star formation, which may either strip away the warm components, leaving cold clouds dominant, or facilitate the conversion of WNM to CNM. In contrast, clouds in the SMC are typically situated in the very outskirts,  where they are more likely influenced by the external warm/hot environment. Further discussion on the differing environments of the SMC and LMC can be found in Section~\ref{sec:dis_env}.

\subsection{Spatial distribution of HI properties}\label{sec:spatialHI}
\begin{figure*}
    \centering
    \includegraphics[width=1\textwidth]{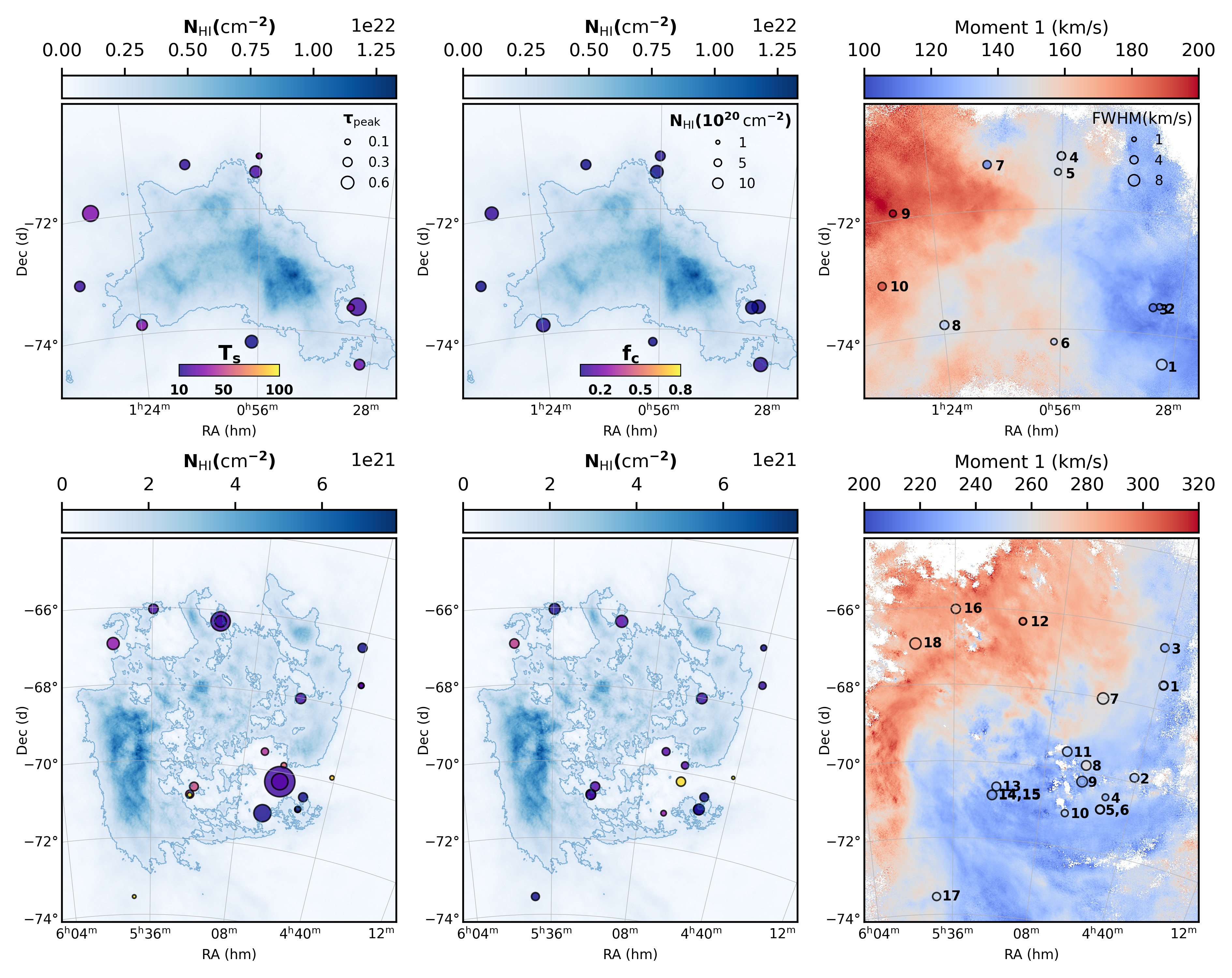}
    \caption{Properties for our selected cold HI structures overplotted on the HI column density map (left and middle columns) and the intensity-weighted velocity field (moment 1) map (right column) of the SMC (upper row) and the LMC (lower row). In the left column, the circles' colors represent the spin temperature of each absorption component, while their sizes indicate the peak optical depth ($\tau_{\text{peak}}$). In the middle column, the circles' colors illustrate the CNM fraction for each line-of-sight, and the sizes reflect the total HI column density from both CNM and WNM for each line-of-sight. Contours in the left and middle column indicate $N_{\text{HI}}\sim 2\times10^{21} ~\text{cm}^{-2}$ for the SMC and $N_{\text{HI}}\sim 1\times10^{21} ~\text{cm}^{-2}$ for the LMC. In the right column, the circles' colors represent the center velocity ($v_{peak}$) of the CNM components, using the same upper and lower limits as the background moment 1 map, while the size corresponds to the FWHM of each absorption component. 
    We only present area in the moment 1 map where $N_{\text{HI}}>10^{19}\text{cm}^{-2}$ for the LMC and $N_{\text{HI}}>10^{20}\text{cm}^{-2}$ for the SMC.}
    \label{fig:property_map}
\end{figure*}

In Figure~\ref{fig:property_map}, we overplot the CNM component and line-of-sight properties on the HI column density and the intensity-weighted velocity field (moment 1) of the SMC and the LMC to examine possible spatial dependencies. 
The circles in the left column represent the spin temperature and peak optical depth for each CNM component, while the circles in the middle column display the CNM fraction and the total HI column density for each line-of-sight. 

In the SMC, there is no special trend for these properties of the cold HI. However, in the LMC, several sources in the southeastern region or the northwestern region, characterized by shell-like structures, show notably high optical depths and high CNM fractions. 
This suggests that shell expansion and interactions with the surrounding medium may be driving cold cloud formation, which we will discuss further in Section~\ref{sec:discuss}.

Circles in the right column of Figure~\ref{fig:property_map} show the central velocities where cold HI is detected, overlaid on the moment 1 map derived from the HI emission data cube. To compare these velocities with the surrounding gas, we calculate the difference between each CNM component’s central velocity and the surrounding gas velocity, which is defined as the average moment 1 velocity within the annulus (two beam - one beam) around each source.  We label the absolute value of this velocity difference as $\Delta V$.
To quantify the effect of turbulent motion of the CNM, we calculate the turbulent velocity of the CNM components \citep[c.f. Equation (16) of][]{Heiles2003b}.
We find an average turbulent velocity of $\sim$ 2 $ \text{km}~ \text{s}^{–1}$ for both the SMC and the LMC (range 1 to 3 km s$^{-1}$).

In the SMC, 2 out of 11 CNM components and in the LMC, 9 out of 23 components have $\Delta V \la 4$ $\text{km}~ \text{s}^{–1}$  \footnote{This value is chosen as the maximum allowed velocity shift between the CNM components seen in emission and absorption during the radiative transfer process.}, reflecting small fluctuations potentially caused by turbulent fluctuations, systematic motions from gravitational collapse and/or stellar activity.
In the SMC, 8 out of 11 CNM components have $\Delta V$ in the range 5 -- 20 $ \text{km}~ \text{s}^{–1}$, which is  typical shell expansion velocity in the SMC \citep{Staveley-Smith1997, Hatzidimitriou2005}. Similarly, 14 out of 23 LMC components have $\Delta V$ in the range 4 -- 25 $ \text{km}~ \text{s}^{–1}$, consistent with the shell expansion velocity in the LMC \citep{Kim1999}. This shows that 
about 18\% of CNM structures in the outskirts of the SMC and about 39\% of CNM structures in the outskirts of the LMC 
have kinematics in agreement with what is expected for turbulent fluctuations or the CNM condensing out of the WNM.
The majority of CNM structures (73\% in the SMC outskirts and 61\% in the LMC outskirts) have kinematics that can be explained by 
shell expansion. We note that in the SMC, because of its long line-of-sight, kinematically different components could be either caused by shell expansion or physically separated structures along the line-of-sight.
Only one SMC structure (Source 7) has $\Delta V \sim$ 40 $ \text{km}~ \text{s}^{–1}$ and is located in the vicinity of  a high-velocity cloud (HVC) identified in \cite{McClure-Griffiths2018} and \cite{Pingel2022}.


\section{Discussion}\label{sec:discuss}

\subsection{Are physical properties of cold HI in the outskirts different from those found inside the SMC/LMC?}

In the SMC, \cite{Jameson2019} reported an average cold gas spin  temperature of $\sim$ 30 K (range from 
$\sim20$ K to $\sim50$ K), which aligns well with our results ($\sim 26$ K). They did not
find any strong trend in the spatial distribution of the spin temperature. As their sources were primarily located within the SMC's main body, our results imply no significant temperature differences between the inner and outer regions of the SMC.

In the LMC, \cite{Liu2021} reported an average cold gas temperature of $\sim$ 28 K (range from $\sim10$ K to $\sim50$ K), similar to that of the SMC. Our results show a slightly higher average cold gas temperature ($\sim40$ K) and a wider distribution for the LMC outskirts. 
Additionally, we find systematically higher spin temperatures for LMC clouds located near shell-like structures. One possibility is that shell motions promote thermal instability and are initiating the early-stage conversion of WNM to CNM. Thus, the CNM is still relatively warm, resulting in higher spin temperatures in these regions. Alternatively, as shells produced by supernovae are found in regions of active star formation, the stellar feedback could be responsible for increased heating, leading to higher spin temperature.

From Figure~\ref{fig:property_CDF_com}, cold HI clouds in the outskirts of the SMC and LMC generally exhibit lower peak optical depths and higher linewidths compared to cold HI clouds within the main bodies of these galaxies. Since the mean spin temperature does not differ much between HI inside and outside the galaxies, the increased linewidths in the outskirts likely result from turbulent broadening.  This is reasonable considering that the majority of cold HI structures have kinematics consistent with systematic motions due to shell expansion. 

We find a lower CNM abundance in the SMC outskirts ($6$ \%) relative to the average CNM fraction of $\sim$ 10 -- 20\% in the SMC \citep{Dempsey2022, Jameson2019}. 
In the LMC outskirts, we find higher CNM fractions than what was found by \cite{Liu2021} ($\sim$ 20 \% vs. $\sim$ 14\%), as well as a very broad distribution. 
The differences in the CNM fraction between the outskirts and the main body are driven by the location of our selected regions. In the SMC, all our selected regions are far from the center and are more exposed to the external environment, such as the Magellanic Corona, rather than being influenced by internal stellar activity in the main body. In contrast, although we refer to the selected regions in the LMC as the ``outskirts'', most are still within the galaxy and near supergiant shells. As a result, internal processes driven by supergiant shell motions have a great impact on the cold gas properties, which results in higher CNM fractions. We will explore this further in the next section.


\subsection{Environments around cold absorbing HI gas}\label{sec:dis_env}
Based on the location of our selected background radio sources, the regions we probe in the SMC are primarily influenced by the external environment, while those in the LMC are dominated by internal processes. In this section, we discuss how these differing environments shape the properties of the cold, absorbing HI.

\subsubsection{The SMC outskirts}
The cold HI in the SMC outskirts exhibits a lower CNM fraction in our samples compared to the SMC main body. Given that our selected regions lie far outside the SMC's main body, they are exposed to the hot Magellanic Corona \citep[$10^{5.3-5.5}$ K;][]{Lucchini2020,Krishnarao2022}, a halo of warm ionized gas centered at the LMC. Small CNM clouds tend to be evaporated to WNM \citep{McKee1977}, with evaporation timescales significantly shorter in hot gas compared to warm gas \citep{Slavin2007}. Consequently, if evaporation occurs in both the outskirts and the main body of the SMC, CNM clouds are harder to survive in the outskirts with a hotter gas environment, leading to the observed low CNM fraction.

\subsubsection{LMC}

In contrast to our SMC sources, most of the regions we study in the LMC are located ``inside'' the main HI disk. Several of these clouds, characterized by high spin temperatures and high CNM fractions, are situated near shells and appear to be influenced by shell motions.

\cite{Kim1999} identified 23 supergiant shells (SGS) and 103 giant shells in the LMC. \cite{Dawson2013} later reorganized these 23 supergiant shells into 11 distinct supergiant shells. 
We correlate the positions of our sources with the supergiant shells listed in \cite{Dawson2013}. Specifically,
Source 12, Source 16, Source 18, Source 7, and Sources 8--11 are associated with SGS5, SGS4, SGS9, SGS6, and SGS7, respectively. These regions also correspond to areas with relatively high CNM fractions in the LMC. 
While molecular gas traced by CO was previously found in these SGSs \citep{Dawson2013}, no CO  has been detected at the position of our sources \citep[based on CO observations from][]{Wong2011}.
\cite{Dawson2013} concluded that molecular cloud formation is enhanced around LMC SGSs, though SGSs are not the dominant driver of molecular gas formation. We find enrichment of CNM around SGSs compared to the CNM in low column density environments. However, whether this conclusion can be applied to the entire LMC requires further analysis about the HI absorption within the whole LMC.

Strong stellar feedback, such as supernova explosions, triggers the formation of these SGSs. In this process, supersonic flows are ejected outward, creating a shock-compressed layer as the flows encounter the surrounding medium. Many simulations have shown that the shock-compressed layers are nonlinearly thermally unstable \citep{Koyama2000}, which promotes a phase transition from WNM to CNM \citep{Audit2005,  Vazquez-Semadeni2006, Audit2010, Kobayashi2020}. This process likely  contributes to the high CNM abundance in regions around SGSs. Additionally, as pointed out by \cite{Clark2023}, those SGSs may more efficiently blow away the less-dense ISM, leaving behind denser ISM with higher dust content. This dense, dust-rich environment could provide shielding that reduces the ambient radiation field, helping to maintain the cold HI gas \citep{McClure-Griffiths2023}.

A few of our selected clouds in the LMC are located in the ``true'' outskirts, including Sources 1--3, and 17. Source 2 is a special case with a 100\% HI being in the CNM phase, which will be further discussed in Section~\ref{sec:source2}.
Sources 1, 3, and 17 have low CNM fractions ($<10$ \%) similar to those found in the SMC outskirts and are also exposed to the hot Magellanic Corona, resulting in the very low CNM fractions.


\subsection{A pure CNM cloud in the LMC}\label{sec:source2}
\begin{figure*}
    \centering
    \includegraphics[width=0.8\textwidth]{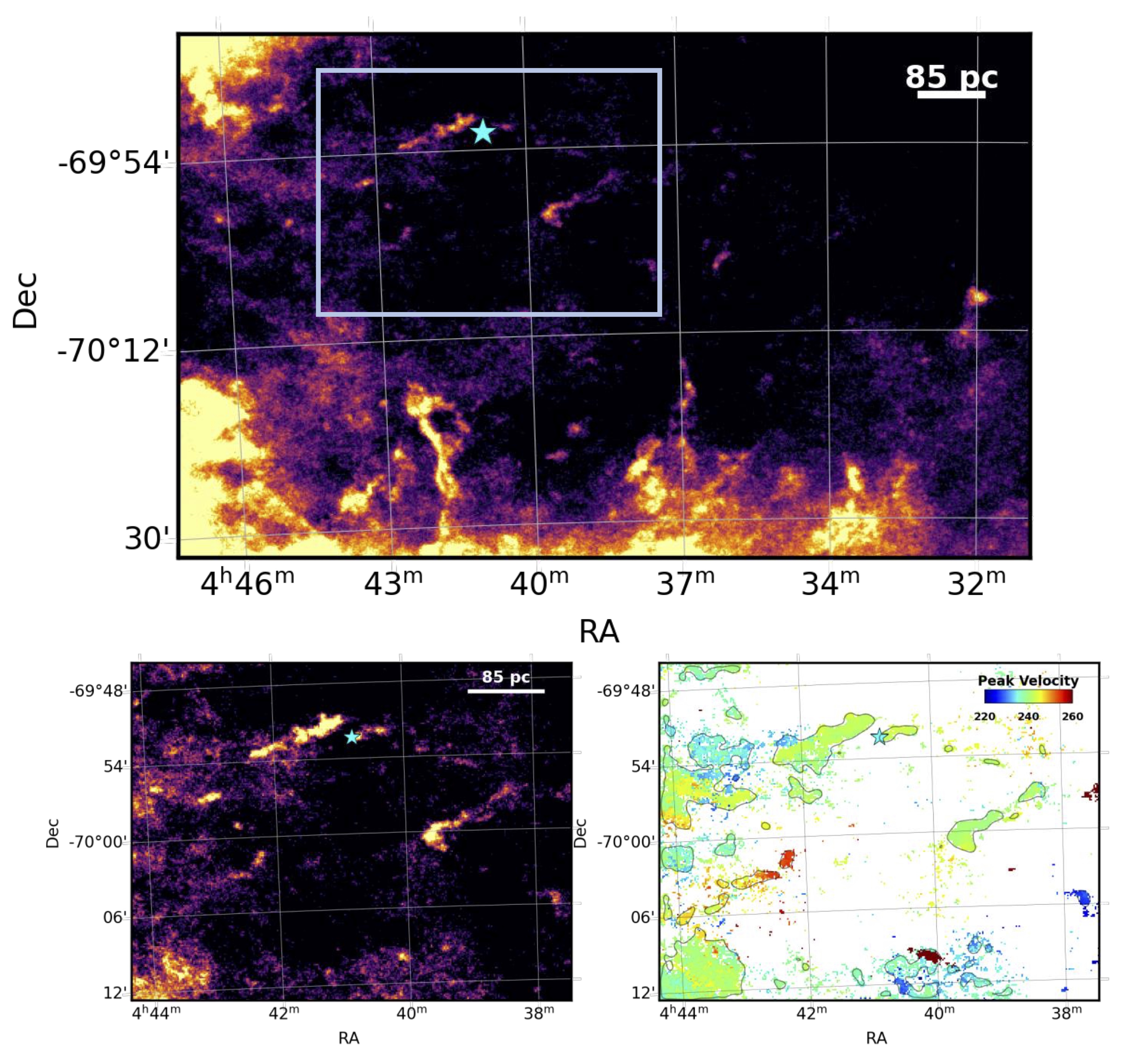}
    \caption{Zoomed-in  view of Source 2 in the LMC. Source 2 is marked as star in these plots. The top panel shows the maximum brightness temperature ($T_{B,max}$) after we smooth each emission spectrum in the datacube by a Gaussian kernel of standard deviation of three channels. The bottom left panel provides a further zoomed-in view of the square region marked in the top panel. The bottom right panel presents the velocity corresponding to the $T_{B,max}$, showing only regions where $T_{B,max}>6$ K.}
    \label{fig:LMC_source2}
\end{figure*}

The HI emission profile for Source 2 can be explained fully with only one velocity component which has $T_s=90$ K and is therefore in the CNM phase.
As a result, this direction which has the lowest HI column density among the LMC directions we probe, has the highest CNM fraction of 1. This is particularly intriguing because in the Milky Way studies like 21-SPONGE (see Figure~\ref{fig:fCNM_NHI}), low column density regions typically exhibit a low
CNM fraction. Moreover, this cloud is located in the very outskirts of the LMC  with no dust content \citep[based on dust observations from][]{Clark2023}, where, similar to other LMC sources and sources around the SMC, 
the influence of the  Magellanic Corona is significant and provides a challenging environment for the survival of cold gas.

In Milky Way HI studies, cold, dense neutral gas is preferentially found in filamentary small-scale structures \citep[e.g.][]{Clark2019, Murray2020,Lei2023}. To explore the morphology of the cloud surrounding Source 2, we zoom in on its surroundings in Figure~\ref{fig:LMC_source2} showing the maximum (peak) brightness temperature distribution. The bottom panel reveals an elongated filamentary structure around Source 2, where the CNM emission profile shown for Source 2 in Figure 8 is consistently present along the filamentary structure, although sometimes a broader and faint component appears nearby. 
We trace the radial velocity of this extended cold HI structure by plotting the radial velocity corresponding to the peak brightness temperature. The bottom right plot shows that the structure has a uniform radial velocity, suggesting a common origin. 
The elliptical morphology of this structure suggests that it could be part of a broken shell. At the location of our observed cold cloud, there is no nearby strong stellar activity and no strong coolants are expected. We speculate that it originated within the LMC and was later expelled by strong stellar feedback. During this process, shocks promoted WNM-to-CNM phase transitions and probably also stripped away much of the WNM, leaving mostly CNM. Simultaneously, it fragmented into small cold clumps and filaments, forming the structure we observe today.

In addition, on an even larger scale, as shown in the top panel, small clumps and filamentary structures extend from the bottom region.  Given that many of these clumps and filaments exhibit emission components with linewidths comparable to or narrower than those of our observed CNM cloud, it is likely that CNM phase dominates in these regions as well. Moreover, these structures have a similar morphology as the supershell walls investigated in \cite{Dawson2011}, who detected molecular gas at the tip of the finger-like supershell walls. 
The survival of the clumpy CNM clouds in this LMC region 
may signal conditions suitable for
the survival of molecular gas.
Future molecular line observations of this region can test this hypothesis.


\subsection{Connection to the Magellanic Stream}
The outskirts of the Magellanic Clouds are connected to the extended outer regions of the Magellanic System.
This system, in addition to its most prominent components -- the SMC and the LMC -- also comprises the Magellanic Bridge connecting the two clouds, the Leading Arm extending ahead of them, and the $\sim 140^{\circ}$ long Magellanic Stream (MS) trailing behind \citep{Nidever2010}. The MS contains low density and low metallicity gas stripped from the Magellanic Clouds, with direct cold HI detection at one location \citep{Matthews2009}. Given the MS's low metallicity and surrounding environment of significant ionized gas, cold HI gas is much harder to survive. Thus, cold HI clouds in the outskirts of the Magellanic Clouds could serve as a reservoir feeding the MS, with shells ejected from the Magellanic Clouds potentially contributing to the supply of cold HI \citep{Dempsey2020}.

\cite{Dempsey2020} found two background radio sources with HI absorption detected in the outskirts of the SMC towards the MS. Their spin temperature estimates via consideration of integrated quantities, along with those from \cite{Matthews2009}, suggested $\sim 70$ K. In our study, Sources 4, 5, 7, 9 in the SMC are located in the direction of the MS. Through radiative transfer fitting, we find $T_s<50$ K.

Additionally, we find that Sources 4, 5, and 7 in the SMC are located near shells identified by \cite{Staveley-Smith1997}, who reported hundreds of shells in the SMC spanning a wide range of scales. These shells, typically circular or elliptical HI structures, are formed by winds from supernovae, massive stars, or gravitational instabilities \citep{Wada2000}.
Source 7, in particular, lies within a broken HI + H$\alpha$ supershell identified by \cite{McClure-Griffiths2018} as being driven by the star formation activity within the galaxy. 
The high-velocity CNM cloud in this direction (see Section~\ref{sec:spatialHI}) is spatially very close to one of the outflowing HI emission clouds \citep[hereafter Alpha cloud, see][]{Buckland-Willis2024} identified by \cite{McClure-Griffiths2018} and \cite{Pingel2022}, with a velocity difference of only a few $ \text{km}~ \text{s}^{–1}$, suggesting that the CNM cloud -- labeled as our Source 7 -- is likely associated with the outflow.

Recent work by \cite{Buckland-Willis2024} analyzed the linewidth of the HI emission across the Alpha cloud and found that it likely carries a significant amount of CNM.
In our study, Figure~\ref{fig:spectra_SMC} shows that the HVC portion of the spectrum (velocity $\sim$ 100 -- 130 $ \text{km}~ \text{s}^{–1}$) has one CNM and one WNM component with the HI column density of $4\times 10^{19}$ cm$^{-2}$ and $1\times 10^{20}$ cm$^{-2}$, respectively. About 30\% of HI associated with the HVC portion of this line-of-sight is in the CNM phase.
The outflowing cold cloud, likely originating from shell expansion,  brings a supply of cold HI to the MS, which supports the conclusion of \cite{Dempsey2020}.

\section{Conclusion}\label{sec:conslu}
We investigate the HI absorption against background radio continuum sources in low column density regions of the SMC and LMC, defined as having an HI column density $N_{\text{HI}}<2\times10^{21} ~\text{cm}^{-2}$ and $N_{\text{HI}}<10^{21} ~\text{cm}^{-2}$ in the SMC and LMC, respectively. Our data, both emission and absorption, are from the recent GASKAP survey, which has the highest spectral and spatial resolution available for the Magellanic Clouds. We apply the radiative transfer method to decompose the emission and absorption spectra, obtaining the individual CNM and WNM components for each source. Following this decomposition, we calculate key properties such as spin temperature, column density, and CNM fraction for each source or component and are able to explore the surrounding environment. Our main findings are summarized below.

\begin{itemize}

\item Our selection criteria resulted in 10 directions with absorbing HI in the outer regions of the SMC. In the LMC, we probe 18 directions with 4 being in the outer regions of the LMC disk and 14 located within the main disk and several are in the vicinity of supergiant shells. 
By comparing the radial velocities of absorbing HI with surrounding HI emission, we find that $\sim$ 20\% of CNM structures in the SMC outskirts and $\sim$ 40\% in the LMC outskirts 
have kinematics in agreement with what is expected for turbulent fluctuations or the CNM condensing out of the WNM.
Most CNM structures (73\% in the SMC and 61\% in the LMC) have kinematics that could be explained by shell expansion, while one cold SMC structure is linked to an outflowing HVC.

\item The selected clouds in the SMC outskirts exhibit low peak optical depths (0.09 -- 1.16) and spin temperatures ($\sim $ 20 K -- 50 K). Their CNM fraction ( 1\% -- 11\%) is significantly lower than that found for the LMC or the Milky Way (e.g. 21-SPONGE), despite having higher total HI column densities. Such low CNM fractions are likely influenced by the SMC's lower metallicity and longer line-of-sight depth, where emission components may be physically separated, with only some of them containing CNM.  
Compared to the SMC's main body \citep{Jameson2019, Dempsey2022}, the outskirts show much lower CNM fractions. This is likely influenced by the exposure of cold HI clouds to the hot Magellanic Corona in the outskirts, where the evaporation of small CNM clouds into WNM occurs much more rapidly.

\item 
Several clouds in the SMC's outskirts likely contribute to the cold HI gas in the Magellanic Stream. The presence of nearby shells suggests that motions induced by these shells could be supplying cold HI to the Stream, consistent with the conclusion of \cite{Dempsey2020}.

\item In the LMC, the selected cold HI clouds show a large range of peak optical depth (0.03 -- 3.55) and spin temperature ($\sim $ 10 K -- 100 K), with no associated CO detections in these CNM clouds.
The CNM fraction has a broad range (1\% -- 100\%) with higher fractions found near supergiant shells identified by \cite{Dawson2013}. 
These shells, driven by strong stellar feedback, produce supersonic flows that can trigger thermal instability, thus facilitating the WNM-to-CNM transition and resulting in high CNM fractions. Enhanced dust shielding around these shells also helps protect the cold HI gas. The thermal instability effect may also explain the pure CNM cloud identified in the far outskirts of the LMC, where a broken shell accompanied by small cold clumps and filaments is found.

\end{itemize}

Our study demonstrates that cold HI in the low column density environments of the SMC and LMC has a lower optical depth and a wider linewidth relative to the HI properties inside the SMC/LMC.
However, a full decomposition of all GASKAP detected HI absorption spectra for the SMC and LMC is essential to enable a fair and comprehensive comparison of cold HI properties between the galaxy outskirts and inner regions in the future.

\section*{Acknowledgments}
We thank the anonymous referee for suggestions that improved the clarity of this paper.
This scientific work uses data obtained from Inyarrimanha Ilgari Bundaran/the Murchison Radio-astronomy Observatory. We acknowledge the Wajarri Yamaji People as the Traditional Owners and native title holders of the Observatory site. CSIRO’s ASKAP radio telescope is part of the Australia Telescope National Facility (\href{https://ror.org/05qajvd42}{https://ror.org/05qajvd42}). 
Operation of ASKAP is funded by the Australian Government with support from the National Collaborative Research Infrastructure Strategy. ASKAP uses the resources of the Pawsey Supercomputing Research Centre. Establishment of ASKAP, Inyarrimanha Ilgari Bundara, the CSIRO Murchison Radioastronomy Observatory and the Pawsey Supercomputing Research Centre are initiatives of the Australian Government, with support from the Government of Western Australia and the Science and Industry Endowment Fund.

This research was partially funded by the Australian Government through an Australian Research Council Australian Laureate Fellowship (project number FL210100039) to NMc-G.
SS acknowledges the support provided by the University of Wisconsin–Madison Office of the Vice Chancellor for Research and Graduate Education with funding from the Wisconsin Alumni Research Foundation, and the NSF Award AST-2108370, and NASA awards 80NSSC21K0991. S.E.C. acknowledges support from NSF award AST-2106607, NASA award 80NSSC23K0972, and an Alfred P. Sloan Research Fellowship.

\section*{Data Availability}
This paper includes archived data obtained through the CSIRO ASKAP Science Data Archive, CASDA (Chapman et al. 2017; Huynh et al. 2020) at \href{https://research.csiro.au/casda}{https://research.csiro.au/casda}.

\appendix
\section{All spectra}\label{sec:all_spec}
\begin{figure}
    \centering
    \includegraphics[width=0.7\textwidth]{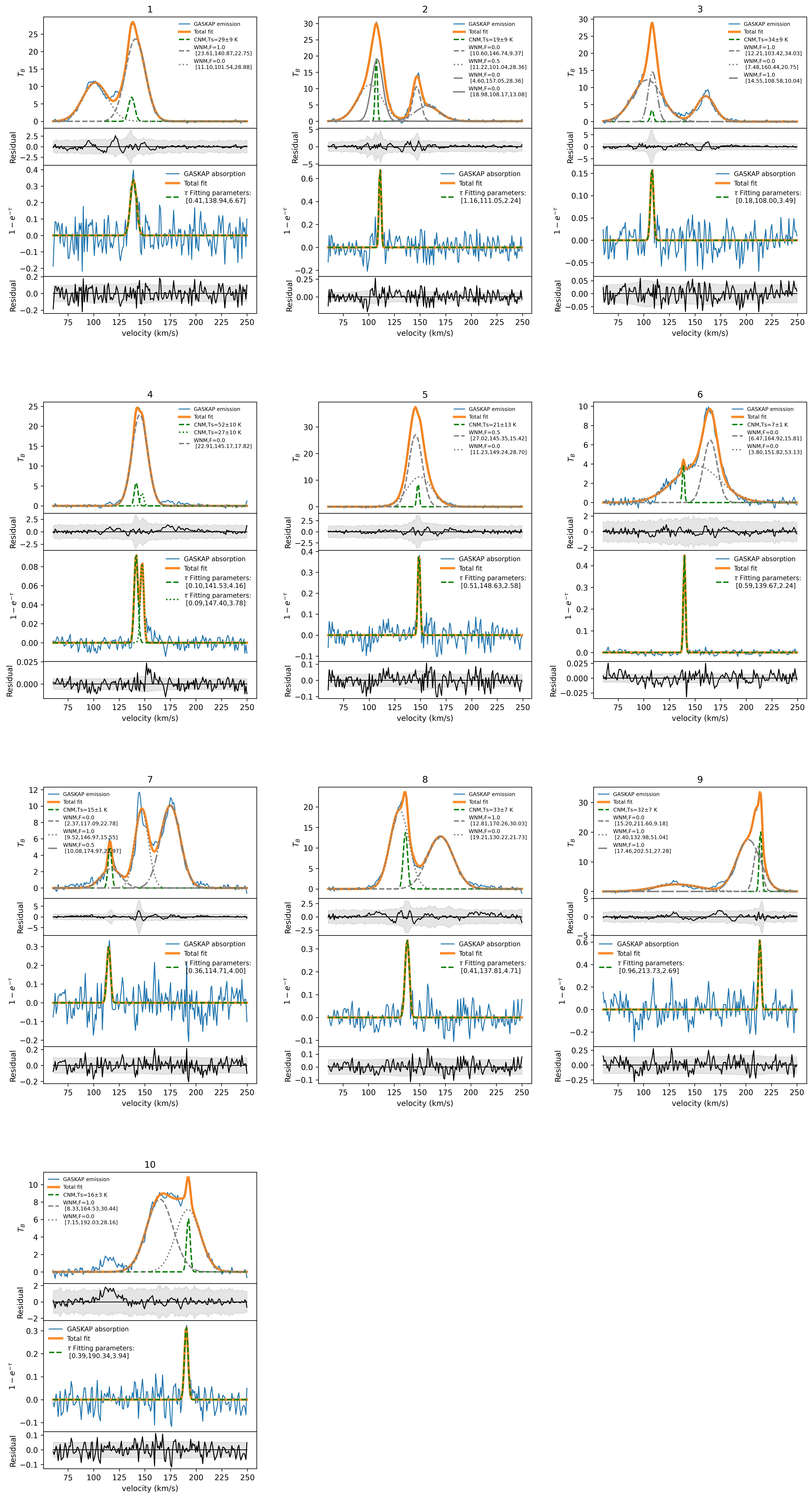}
    \caption{All fitting results of the SMC emission and absorption spectra described in Section~\ref{sec:decomp}. Plot for each source has same configuration as Figure~\ref{fig:spectra_single_SMC}.}
    \label{fig:spectra_SMC}
\end{figure}

\begin{figure}
    \centering
    \includegraphics[width=0.7\textwidth]{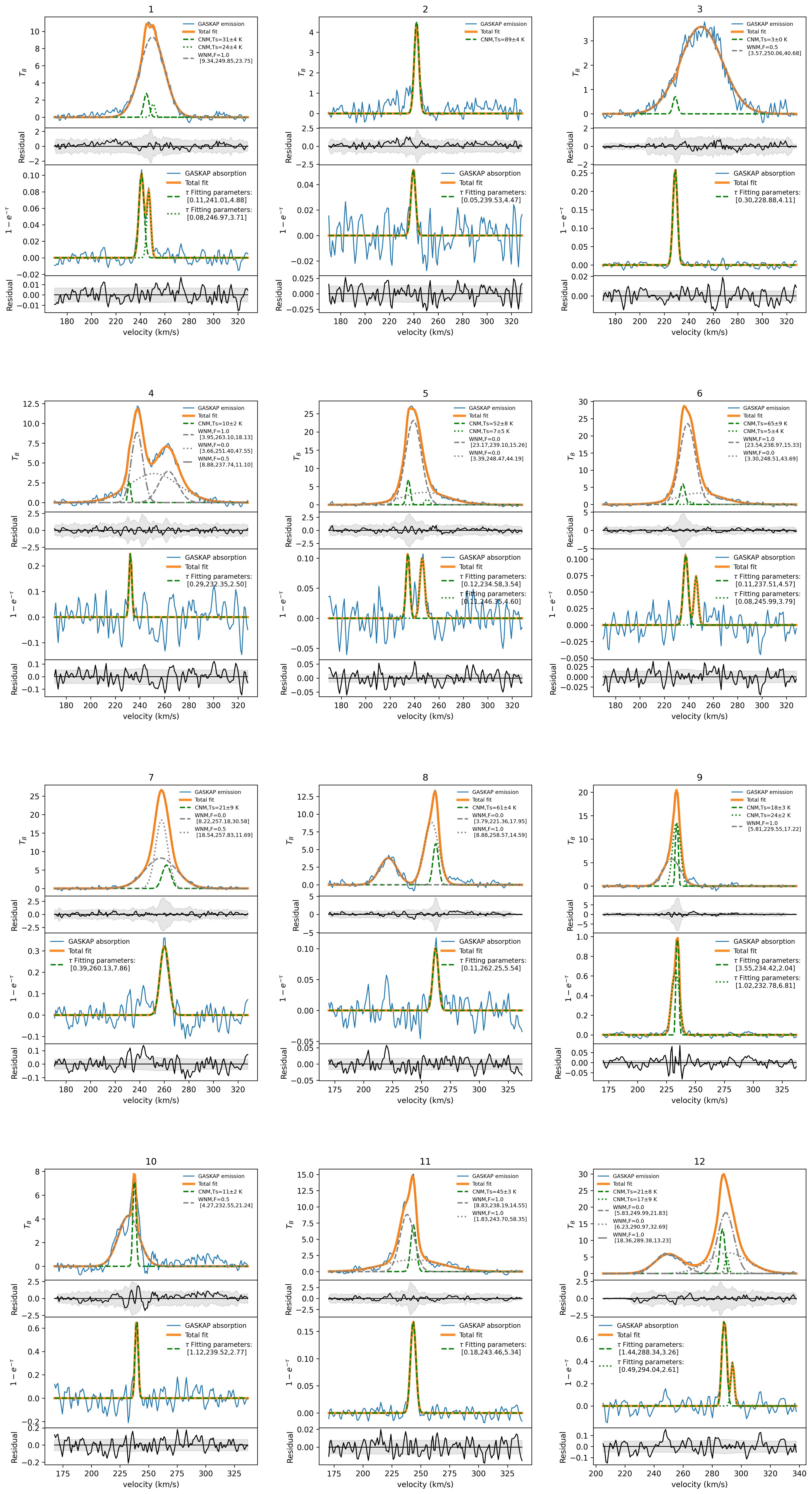}
    \caption{All fitting results of the LMC emission and absorption spectra described in Section~\ref{sec:decomp}. Plot for each source has same configuration as Figure~\ref{fig:spectra_single_SMC}.}
    \label{fig:spectra_LMC1}
\end{figure}

\begin{figure}
    \centering
    \includegraphics[width=0.7\textwidth]{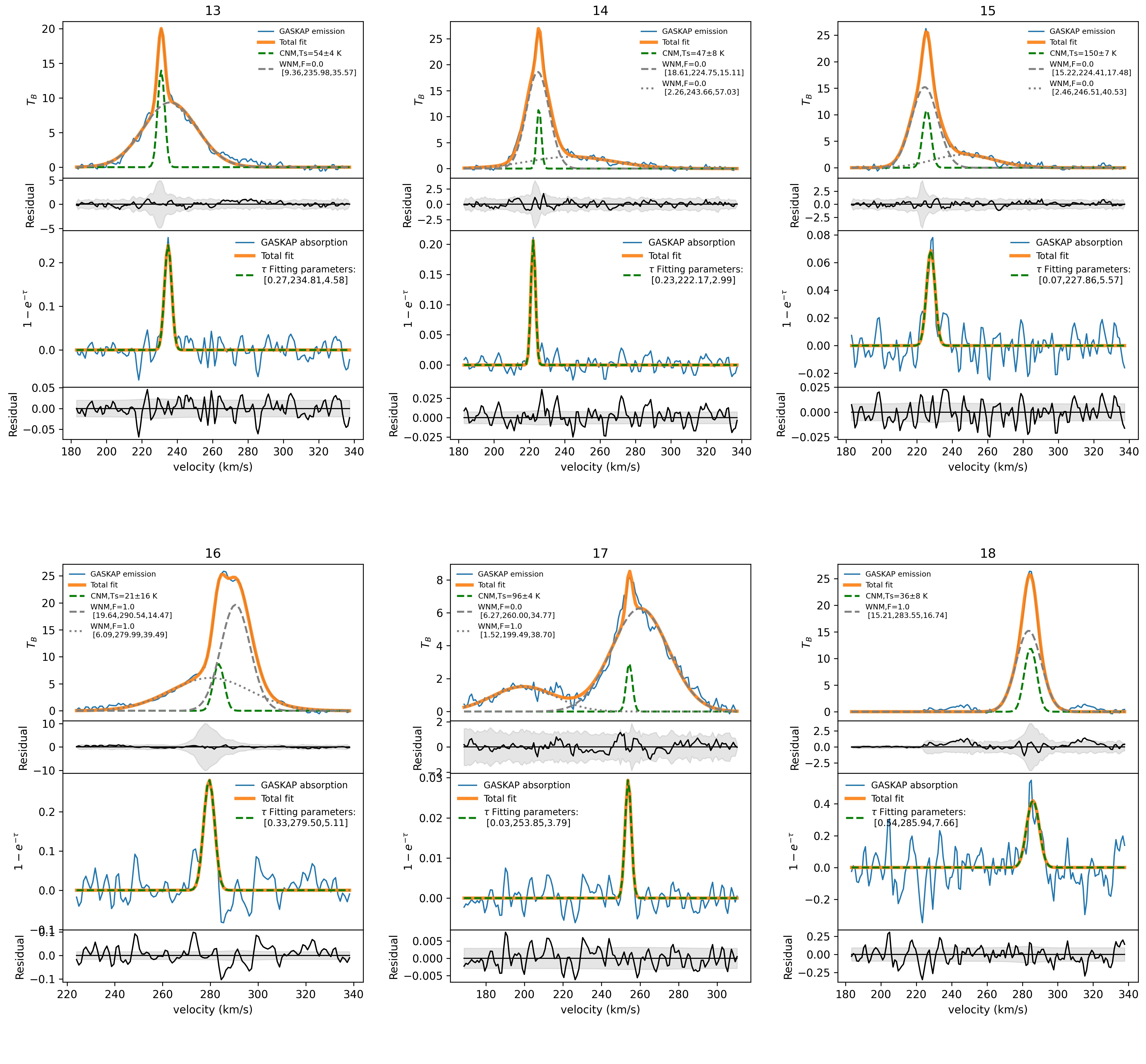}
    \caption{Figure~\ref{fig:spectra_LMC1} continued.}
    \label{fig:spectra_LMC2}
\end{figure}

In this appendix, we include the Gaussian decomposition fitting results of all sources in the SMC and the LMC, as shown in Figure~\ref{fig:spectra_SMC}, Figure~\ref{fig:spectra_LMC1} and Figure~\ref{fig:spectra_LMC2}.

\bibliography{sample631}{}
\bibliographystyle{aasjournal}

\end{document}